\documentclass[fleqn,usenatbib]{mnras}

\usepackage{newtxtext,newtxmath}

\usepackage[T1]{fontenc}
\usepackage[normalem]{ulem}

\DeclareRobustCommand{\VAN}[3]{#2}
\let\VANthebibliography\thebibliography
\def\thebibliography{\DeclareRobustCommand{\VAN}[3]{##3}\VANthebibliography}

\usepackage{graphicx}	
\usepackage{amsmath}	
\usepackage{multirow}
\usepackage[dvipsnames]{xcolor}

\title[Cosmic evolution of Active Galactic Nuclei]{Cosmic evolution of  the incidence of Active Galactic Nuclei in massive clusters:
Simulations versus observations}

\author[I. Mu\~noz-Rodr\'iguez et al.]{
Iv\'an Mu\~noz Rodr\'iguez$^{1,2}$\thanks{E-mail: ivan.rodriguez@noa.gr},
Antonis Georgakakis$^{1}$,
Francesco Shankar$^{2}$,
Viola Allevato$^{3}$,
Silvia Bonoli$^{4,5}$,
\newauthor Marcella Brusa$^{6,7}$,
Andrea Lapi$^{8}$ and
Akke Viitanen$^{9}$
\\
$^{1}$Institute for Astronomy and Astrophysics, National Observatory of Athens, V. Paulou \& I. Metaxa 11532, Greece\\
$^{2}$Department of Physics and Astronomy, University of Southampton, High-field SO17 1BJ, UK\\
$^{3}$INAF - Astronomical Observatory of Capodimonte, Salita Moiariello 16, 80131 Naples, Italy\\
$^{4}$Donostia International Physics Center (DIPC), Manuel Lardizabal Ibilbidea, 4, San Sebastián, Spain\\
$^{5}$Ikerbasque, Basque Foundation for Science, E-48013 Bilbao, Spain\\
$^{6}$INAF - Osservatorio di Astrofisica e Scienza dello Spazio di Bologna, via Gobetti 93/3, 40129 Bologna, Italy\\
$^{7}$Dipartimento di Fisica e Astronomia "Augusto Righi", Universit\`a di Bologna, via Gobetti 93/2, 40129 Bologna, Italy\\
$^{8}$SISSA, Via Bonomea 265, 34136 Trieste, Italy\\
$^{9}$Department of Physics, Gustaf H\"allstr\"omin katu 2, 00014 University of Helsinki, Finland
}

\date{Accepted XXX. Received YYY; in original form ZZZ}

\pubyear{2022}

\begin{document}
\label{firstpage}
\pagerange{\pageref{firstpage}--\pageref{lastpage}}
\maketitle
\begin{abstract}
This paper explores the role of small-scale environment ($<1$\,Mpc) in modulating accretion events onto supermassive black holes by studying the incidence of Active Galactic Nuclei  (AGN) in massive clusters of galaxies. A flexible, data-driven semi-empirical model is developed based on a minimal set of parameters and under the zero order assumption that the incidence of AGN in galaxies is independent of environment. This is used to predict how the fraction of X-ray selected AGN among galaxies in massive dark matter halos ($\ga3\times 10^{14}\,M_{\odot}$) evolves with redshift and reveal tensions with observations. At high redshift, $z\sim1.2$, the model underpredicts AGN fractions, particularly at high X-ray luminosities, $L_X(\rm 2-10\,keV) \ga 10^{44}\, erg \, s^{-1}$. At low redshift, $z\sim0.2$, the model estimates fractions of moderate luminosity AGN ($L_X(\rm 2-10\,keV) \ga 10^{43}\, erg \, s^{-1}$) that are a factor of $2-3$ higher than the observations. These findings  reject the zero order assumption on which the semi-empirical model hinges and point to a strong and redshift-dependent influence of the small-scale environment on the growth of black holes. Cluster of galaxies appear to promote AGN activity relative to the model expectation at $z\sim1.2$  and suppress it close to the present day. These trends could be explained by the increasing gas content of galaxies toward higher redshift combined with an efficient triggering of AGN at earlier times in galaxies that fall onto clusters.

\end{abstract}

\begin{keywords}
galaxies: haloes -- galaxies: active -- galaxies: clusters: general -- galaxies: nuclei -- quasars: supermassive black holes -- galaxies: Seyfert -- X-rays: galaxies: clusters
\end{keywords}



%
%
%
%
%

\section{Introduction}
It is now widely accepted that supermassive black holes are found at the centres of most, if not all, galaxies in the local Universe \citep[][]{Kormendy_Ho2013}. These compact objects are believed to have grown their masses throughout cosmic time via accretion of material from their surroundings \citep[e.g.][]{Soltan1982, Marconi2004, Merloni_Heinz2008}. Such accretion events generate large amounts of energy that can be detected as radiation across the electromagnetic spectrum. The astrophysical sources associated with such events are dubbed Active Galactic Nuclei \citep[AGN, ][]{Padovani2017}. Observational campaigns in the last 20 years aiming at detecting and characterising large samples of AGN have painted a comprehensive picture of the cosmological evolution of this population and have provided a quantitative description of the accretion history of the Universe out to high redshift \citep[e.g.][]{Ueda2014, aird15_xlf, Brandt_Alexander2015}. What remains challenging to understand however, are the physical processes that trigger accretion events onto supermassive black holes and therefore drive the observed black hole growth as a function of redshift. The different supermassive black hole fueling mechanisms proposed in the literature  can broadly be grouped into external ({\it ex-situ}) and internal ({\it in-situ}) in nature. The latter are related to the secular evolution of galaxies, e.g. disk instabilities \citep[e.g.][]{Hopkins_Hernquist2006, Gatti2016}, the creation of bars \citep{Cisternas2015}, stellar winds \citep{Ciotti_Ostriker2007, Kauffmann_Heckman2009} or the biased collapse of the baryons in the inner region of the halo \citep[e.g.][]{Lapi2011, Lapi2018}, and could lead to gas inflows toward the central regions of the galaxy and feeding of the central black hole. {\it Ex-situ} processes are those that act on a galaxy from its environment. They include for example, galaxy interactions \citep{DiMatteo2005, Gatti2016}, cold gas inflows \citep{Bournaud2012, DeGraf2017}, or cooling flows in massive clusters \citep[e.g.][]{Fabian1994}. The balance between {\it ex-situ} and {\it in-situ} supermassive black hole fuelling processes likely depends, among others, on redshift, position or the cosmic web and intrinsic galaxy properties, such as gas content and structural parameters.

One approach for exploring the relative importance of the diverse mechanisms above in modulating the growth of supermassive black holes is to study the incidence of AGN in galaxies as a function of e.g. their morphology, star-formation rate (SFR) or position on the cosmic web \citep{Brandt_Alexander2015}. Such investigations can shed light on the conditions that promote or suppress accretion events onto the supermassive black holes of galaxies and make inferences on the physics at play. Environmental studies in particular, i.e. how AGN populate galaxy groups, filaments, clusters and field, could provide information on the balance between {\it in-situ} and {\it ex-situ} process for activating supermassive black holes. This potential have motivated observational studies to characterise AGN populations in different environments. At low redshift ($z\approx0.1$) there is evidence that the fraction of AGN in high density regions is lower compared to the field \citep[e.g.][]{Kauffmann2004, Popesso_Biviano2006, Koulouridis2010, Lopes2017, Mishra_Dai2020}. This may indicate the decreasing incidence of mergers in massive clusters \citep{Popesso_Biviano2006} and/or the impact of processes that strip the gas reservoirs of galaxies and hence, lead to the suppression of their nuclear activity. There are also claims that the AGN radial distribution is skewed to the cluster outskirts relative to the general galaxy population \citep[e.g][]{deSouza2016, Lopes2017}. This finding coupled with suggestions that cluster AGN show high velocity dispersion \citep[e.g][]{Haines2012, Pimbblet2013, Lopes2017} points to a link between accretion events and galaxies that fall onto high density regions from larger scales. Contrary to the findings above there are also observational studies that claim little or no dependence of the AGN fraction on environment at low redshift \citep[e.g.][]{Miller2003, Haggard2010, Pimbblet2013}. At least part of the discrepancy is likely related to selection effects. These include the accretion luminosity threshold adopted in the various studies \citep{Kauffmann2004, Pimbblet2013}, differences between field and cluster environments in the properties of the overall galaxy population (e.g. SFR, morphology) used to determine fractions \citep[e.g.][]{vonderLinden2010, Lopes2017, Man2019} or the methods adopted for selecting AGN (e.g. optical emission lines, X-ray emission, mid-infrared colours). 

Outside the local Universe ($z\ga0.1$) there is evidence that the group/cluster environments become more active in terms of black hole growth. The fraction of AGN in such dense regions increases with increasing redshift \citep{martini09, martini13, Bufanda2017} at a rate that appears to be faster than the field AGN evolution \citep{Eastman2007}. At redshift $z\ga1$ the fraction of AGN in clusters is at least as high as the field expectation \citep{martini13, alberts16} suggesting efficient triggering of accretion events. This is possibly associated with the higher incidence of interactions in these environments \citep[e.g.][]{alberts16} and/or the larger cold gas content of galaxies at earlier times \citep[e.g.][]{Tacconi2010} combined with the impact of the ram-pressure experienced by galaxies as they fall into the cluster potential well \citep[][]{Poggianti2017, Ricarte2020}. The evidence above emphasises the role of environment for understanding AGN triggering mechanisms and underlines the need to better constrain the redshift at which the cluster AGN fractions are on par with the field or even exceed it \citep[][]{alberts16}.

In this work we revisit the incidence of AGN in massive clusters of galaxies out to $z\approx1.25$ by developing a semi-empirical modeling approach to interpret observational results from the literature \citep{martini09,martini13}. The feature of our modeling methodology is control over systematics and observational selection effects. We use observationally motivated relations to populate massive dark matter halos extracted from N-body cosmological simulations with AGN and galaxies. These are then used to mimic observations of clusters of galaxies by including in a realistic manner the relevant selection effects, such as cluster membership definition, flux or luminosity cuts, etc. These mocks are then compared with real observations to make inferences on the evolution of the AGN fraction in clusters relative to the field expectation.  Section~\ref{sec:obs} presents the observations and selection bias that we attempt to reproduce. Section~\ref{sec:methods} describes the generation of the mock catalogues and the implementation of the different selection effects into the simulations. The comparison of the semi-empirical model predictions with the observations is presented in Section~\ref{sec:results}. Finally, Section~\ref{sec:discussion} discusses the results in the context of AGN triggering mechanisms. We adopt a flat $\Lambda$CDM cosmology with parameters $\Omega_{\rm{m}}$ = 0.307, $\Omega_{\rm{\Lambda}}$= 0.693, $h = 0.678$ consistent with the Planck results \citep{planck16}.

\section{Observations}\label{sec:obs}
This work uses the observational measurements of the fraction of X-ray AGN in galaxy clusters presented by \cite{martini09,martini13}. Typical halo masses of these clusters are few times $10^{14}\,M_\odot$ (see Section~\ref{sec:clust_sample} for more details). In this section we describe the most salient details of these observations and the corresponding data analysis. Of particular interest to our work are the (i) the definition of cluster membership in the observations and (ii) the magnitude/flux limits that are used to define the galaxy and AGN samples. The inferred AGN fractions strongly depend on these selection effects and it is therefore important to reproduce them in the simulations before comparing with the observations.

\cite{martini09} used a sample of 32 massive galaxy clusters out to redshift $z=1.3$ with available \textit{Chandra} X-ray observations. Their low redshift sub-sample consists of 17 clusters at  $z<0.4$ (mean redshift $\bar{z}=0.19$). These 17 clusters include the 10 presented in \cite{martini06} and 7 additional ones selected from the \textit{Chandra} archive to be the nearest most massive clusters with virial radius that fits within the \textit{Chandra} field-of-view (FOV). The high redshift sub-sample numbers 15 clusters in the redshift interval $z=0.4-1.3$ (average redshift $\bar{z}=0.72$). Cluster member candidates are selected within the projected $R_{200}$\footnote{The virial radius of a cluster is defined to be the distance from the cluster center where the local density is 200 times the mean density of the Universe.} radius of each cluster in the sample. The number of AGN and galaxy cluster members is determined to the apparent $R$-band magnitude limit $m_{R}^{*}(z)+1$, where $m_{R}^{*}(z)$ is the break of the $R$-band luminosity function at the cluster redshift. The latter is estimated assuming that the absolute magnitude break of the luminosity function evolves as $M_R^*(z) = M_R^*(z=0)-z$  with $M_R^*(z=0)$ from \cite{christlein03_break_lf} and early type galaxy spectral energy distribution for the K-correction. For clusters with a high redshift identification completeness the number of galaxy members is estimated by counting sources with $R$-band magnitude brighter than $m_R^*(z)+1$ and redshift difference ($\Delta z$) relative to the mean cluster redshift ($z$), $\Delta z \cdot c < 3\,\sigma_v (1+z)$, where $\sigma_v$ is the cluster velocity dispersion and $c$ the  speed of light. For clusters with limited spectroscopic redshift follow-ups (mostly high-redshift sub-sample) the number of galaxy members is estimated using the cluster-richness vs velocity dispersion relation of \cite{becker07_richnes_sigma}. This empirical relation is calibrated to yield the number of early-type galaxy cluster members that are more luminous than 0.4\,$L^*$ (i.e. equivalent to $m_R^*(z)+1$) within the $R_{200}$ radius. AGN cluster members are also selected to have apparent magnitude brighter than $m_R^*(z)+1$ and redshifts that are consistent with $\Delta z\cdot c < 3\,\sigma_v(1+z)$, i.e. similar to galaxies. The observed number of X-ray AGN cluster members is also corrected for the spectroscopic completeness of each cluster (typically $>60$\% for AGN). The depth of the {\it Chandra} X-ray observations means that AGN samples are complete to hard-band luminosities $L_X(\rm 2-10\,keV) >10^{43}\, erg\,s^{-1}$. Less luminous X-ray sources suffer incompleteness because of the sensitivity of the \textit{Chandra} observations and are not used for the estimation of AGN fractions.

The \cite{martini13} cluster sample is composed of 13 of the most statistically significant extended X-ray sources detected in the {\it Chandra} survey of the Bootes field with spectroscopic identifications  in the redshift interval $z= 1-1.5$ \citep{Eisen08_clust_selec}. Cluster member candidates, AGN or galaxies, are selected to lie within the projected $R_{200}$ radius and have {\it Spitzer} $3.6\,\mu \rm m$-band apparent magnitude brighter than $m_{3.6}^*(z)+1$. The quantity $m_{3.6}^*(z)$ is the break of the $3.6\,\mu \rm m$  luminosity function at redshift $z$ adopted from \cite{mancone10}. Both spectroscopic and photometric redshifts are used to determine cluster membership. Sources with spectroscopic redshift within $\Delta z\cdot c<\pm2\,000 (1+z)\,\rm km\,s^{-1}$ (i.e. similar to \citealt{martini09} fixing $3\sigma_v=2\,000\,\rm km\,s^{-1}$) off the cluster redshift are assumed to be members. In the case of photometric redshift estimates this condition is modified so that at least 30\% of the photometric redshift probability density function is required to lie within the above redshift interval. All X-ray selected AGN cluster member candidates in the sample of \cite{martini13} have spectroscopic redshifts. The AGN sample is complete to the X-ray luminosity $L_X(\rm 2-10\,keV) =10^{44}\, erg \, s^{-1}$. For less luminous systems, $L_X(\rm 2-10\,keV) >10^{43}\, erg \, s^{-1}$, \cite{martini13} provide lower limits for the AGN cluster fraction. 

\section{Methodology}\label{sec:methods}
This section describes our approach for generating mock observations of galaxies and AGN in massive structures of the cosmic web. The starting point of our method are cosmological N-body simulations \citep[e.g. ][]{millennium2006, bolshoi2011, MultiDark2016} that describe the formation and evolution of dark matter halos in the Universe under the influence of gravity. These are coupled with empirical relations that associate dark matter halos with galaxies (galaxy-halo connection). Accretion events associated to supermassive black holes are then painted on top of those galaxies using recent observational results on the incidence of AGN in galaxies (AGN-galaxy connection). The implicit assumption of this latter step is that the probability of galaxies hosting an accretion event does not depend on environment, i.e.,  halo mass. Light-cones are then generated to mimic real observations of AGNs and galaxies on the cosmic web. These steps above are described in detail in the following sections.

\subsection{Galaxy-halo connection: (Sub-)Halo Abundance matching techniques}

It is well established that the main sites of galaxy formation in the Universe are halos of dark matter. These provide the necessary gravitational potential for the various baryonic physical processes to act and form the luminous structures (i.e. galaxies) we observe. Among the different methods proposed in the literature for associating galaxies (i.e. luminous baryonic matter) with dark matter halos, the semi-empirical approach of abundance matching offers a number of advantages. With relatively small number of parameters, this approach can successfully reproduce observed properties of galaxies such as their stellar masses or SFRs. In the basic implementation of abundance matching it is assumed that most massive halos are associated with the most massive galaxies. This approach yields a relation between dark matter mass and stellar mass as a function of redshift that is in reasonable agreement with observational results \citep[e.g. occupation number, two point correlation function or cross bias, see ][]{vale04_am,kravtsov04_am,tasitsiomi04_am}. Recent implementations of abundance matching techniques include an increasing level of complexity in the way halos are associated with baryonic mass and galaxies. For example, the halo mass vs. stellar mass relation is parameterized by analytic functions allowing for intrinsic scatter \citep[e.g. ][] {behroozi10_scatter, moster10_scatter}, baryonic process such as star-formation in galaxies are modeled using information on the accretion/merger history of halos, diverse observational results (e.g. stellar mass functions, galaxy clustering properties) are used to tune the various model parameters and produce realistic mock galaxy catalogues out to high redshift \citep{behroozi19_um, moster18_emerge}.

In this work we use the \textsc{UniverseMachine} data release 1\footnote{\url{https://www.peterbehroozi.com/data.html}} \citep[][]{behroozi19_um} implemented for the MultiDark PLanck2 \citep[MDPL2, ][]{MultiDark2016} dark matter N-body simulation. We choose to use the MDPL2 because it is one of the largest volume, high resolution and public cosmological simulations. It has a box size of 1\,000 Mpc/$h$, a mass resolution of $1.5\times10^9\,M_\odot/h$ and 3\,840$^3$ ($\sim$57$\cdot$10$^9$) particles. Individual dark matter halos in the MDPL2 are identified using \textsc{Rockstar} \citep{behroozi13_rockstar}. This is a state-of-the-art halo finder that uses both the 6-dimensional phase-space distribution of dark matter particles and temporal information to identify bound structures, i.e. dark matter halos.
\textsc{Rockstar} is efficient in detecting and measuring the properties of both the largest collapsed structures (parent haloes) and sub-structures within them (satellites haloes). The evolution of haloes through cosmic time is tracked in the form of merger trees computed by the code \textsc{Consistent Trees} \citep{behroozi13_consist_trees}. In this work we consider only dark matter halos with at least 100 times the MDPL2 mass resolution, i.e. $M_{\rm peak}>1.5\times10^{11}\,M_\odot/h$. This limit ensures that the inferred properties of dark matter halos, such as their position and total mass are not affected by the finite resolution of the simulations.

\textsc{UniverseMachine} assigns stellar masses (and hence galaxies) to dark matter halos by parameterizing the star formation history (SFH) of individual halos. The SFR in a halo is assumed to be a function of the depth of the halo's potential well, its assembly history and cosmic time. The maximum circular velocity, $v_{\rm max}$, is used as a proxy of the depth of the potential well. The $v_{\rm max}$, corresponds to the circular velocity of the halo when it reaches its historical maximum mass ($M_{\rm peak}$ parameter in the MDPL2 catalogues). The halo assembly history is parametrized by the $v_{\rm max}$ variations ($\Delta v_{\rm max}$) across cosmic time. \textsc{UniverseMachine} therefore assumes a parametric analytic function SFR$(v_{\rm max}, \Delta v_{\rm max}, z)$ to determine the SFR for each halo across cosmic time. Integrating the SFR along the assembly and merger history of a galaxy it is then possible to determine the corresponding stellar mass. The parameter space of SFR$(v_{\rm max}, \Delta v_{\rm max}, z)$ function is explored in an iterative manner by estimating at each step observables (stellar mass functions, UV luminosity functions, the UV–stellar mass relation, specific and cosmic SFRs, galaxy quenched fractions, galaxy auto-correlation functions and the quenched fraction of central galaxies as a function of environmental density) and comparing them with observations at different redshfits. A Monte Carlo Markov Chain (MCMC) approach is used to sample the model parameter space and yield posteriors for the model parameters.

The end product of \textsc{UniverseMachine} are catalogues of dark matter halos, each of which is assigned a galaxy stellar mass and a SFR. By construction the galaxy population is consistent with observations, including the stellar mass function at different redshift, the evolution of the SFR density of the Universe and the Main Sequence of star-formation. In the following we use the ``observed'' \textsc{UniverseMachine} values for the stellar mass and SFR of mock galaxies. These are estimated by adding systematic errors (also free parameters in \textsc{UniverseMachine}) to the ``true'' values to account for observational effects (e.g. Eddington bias). We note however, that our final results and conclusions are not sensitive to this choice. For dark matter haloes we use virial values as defined by \cite{bryan98_deltavir} for mass and radius.

\subsection{AGN-galaxy connection: specific accretion rate distributions} \label{sec:agn_galax_conec}

The assignment of AGN to the \textsc{UniverseMachine} galaxies is also based on empirical relations that associate the probability of a supermassive black hole accretion event to the properties of its host. The relevant observable is the specific accretion rate, $\lambda_{\rm{sBHAR}} \propto L_X({\rm 2-10\,keV}) /M_\star$. In this definition $L_X(\rm 2-10\, keV)$ is the AGN X-ray luminosity in the $2-10$\,keV band and $M_\star$ is the stellar mass of the parent galaxy. The specific accretion rate provides an estimate of how much X-ray luminosity is emitted by the AGN per unit stellar mass of the host galaxy. In this work we choose to scale the ratio $L_X({\rm 2-10\, keV})/M_\star$ as 

\begin{equation}
    \lambda_{\rm{sBHAR}} = \frac{\rm{k}_{\rm{bol}}}{1.26 \times 10^{38} \times 0.002}  \cdot \left(\frac{L_X(2-10\,\rm{keV})}{\rm{erg}\,\rm{s}^{-1}}\right) \cdot \left( \frac{M_\odot}{M_\star}\right).
    \label{eq:sar_def}
\end{equation}

\noindent The above equation assumes a Margorrian-type relation between stellar and black hole mass $\rm{M}_{BH} = 0.002 \cdot M_\star$ \citep[][]{marconi03}, an AGN bolometric correction $\rm{k}_{\rm{bol}} = L_{\rm{bol}} / L_X(2-10\,\rm{keV}) = 25$ \citep[][]{elvis94} and the Eddington luminosity of the black hole $1.26\times10^{38}\,\rm{erg\,s^{-1}}$. The scaling factors in Equation~(\ref{eq:sar_def}) make $\lambda_{\rm{sBHAR}}$ resemble an Eddington ratio, i.e. the AGN bolometric luminosity normalised to the Eddington luminosity of the black hole. It is emphasised that the multiplicative constants in Equation~(\ref{eq:sar_def}) do not affect our analysis and the assignment of AGN luminosities to \textsc{UniverseMachine} galaxies.

Large multi-wavelength observational programs have enabled the estimation of stellar masses, X-ray luminosities and hence $\lambda_{\rm{sBHAR}}$ for large samples of AGN \citep{aird12_primus, bongiorno12_cosmos, schulze15, age17_sar}. These observations made possible the determination of the fraction of galaxies at fixed stellar mass that host an accretion event with specific accretion rate $\lambda_{\rm{sBHAR}}$. These fractions can then be turned into specific accretion rate probability distribution functions, P($\lambda_{\rm{sBHAR}}$), which describe the probability of an accretion event with parameter $\lambda_{\rm{sBHAR}}$ in a galaxy. Recent observational studies have measured the specific accretion rate distribution as a function of redshift and host galaxy properties such as stellar mass and SFR \citep[][]{aird18_sar,age17_sar}. In this work we use these two independent estimates of the specific accretion rate distribution. 

\begin{figure}
    \centering 
    \includegraphics[width=0.45\textwidth]{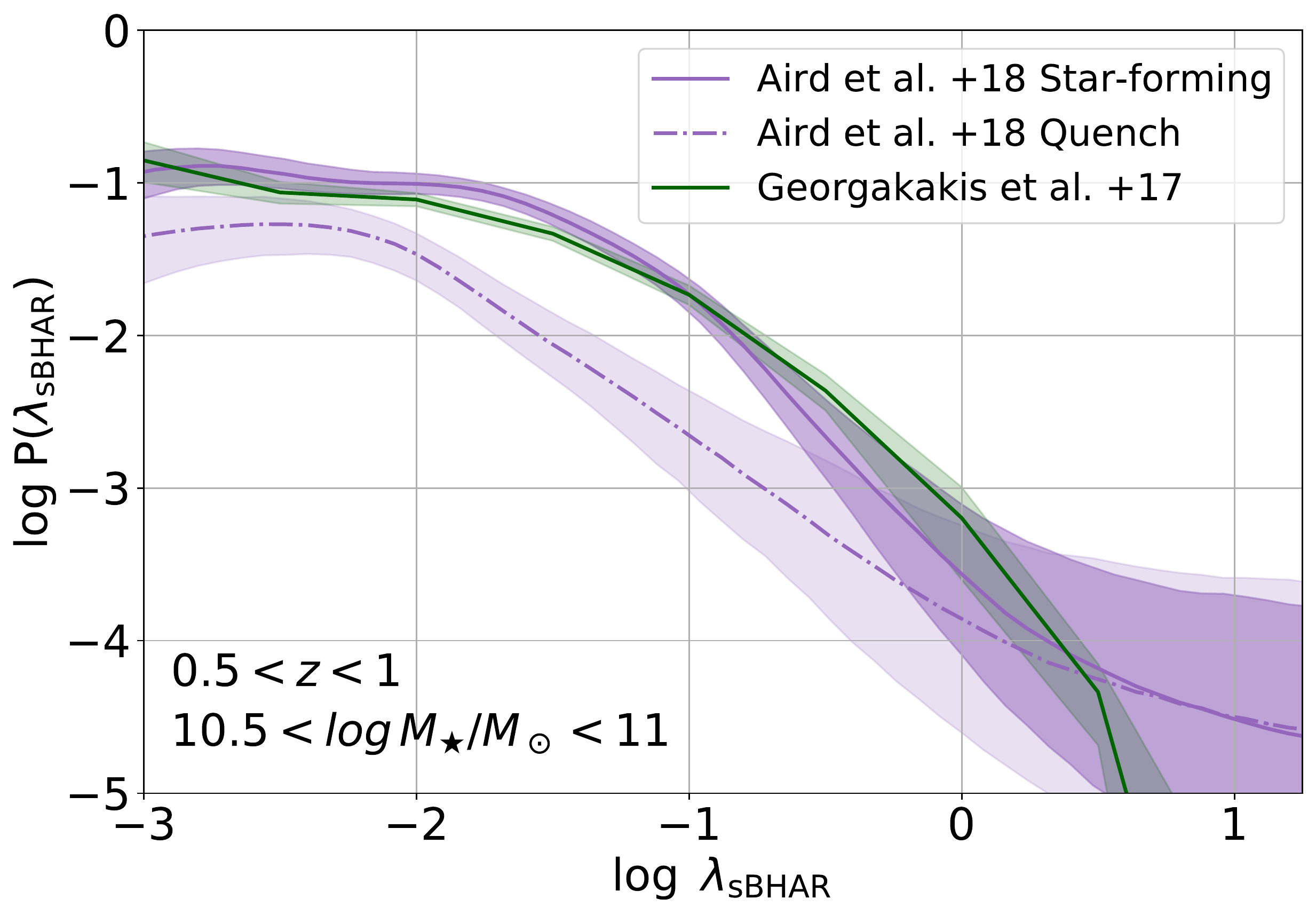}
    \caption{Specific accretion rate distributions that describe the probability of a galaxy hosting an AGN with specific accretion rate $\lambda_{\rm sBHAR}$. The shaded regions correspond to different observational measurements of P($\lambda_{\rm sBHAR}$). The purple colour shows the \citet{aird18_sar} result, where different line styles indicate different galaxy types: star-forming galaxies (solid line) or passive (dash-dotted). The green colour shows the \citet{age17_sar} constraints on the specific accretion are distribution. All curves correspond to the redshift interval $z=0.5-1$ and stellar mass interval $M_\star=10^{10.5}-10^{11}M_\odot$. The extent of the shaded regions correspond to the 68\% confidence interval around the median (bold curves).}
    \label{fig:sar}
\end{figure}

\cite{age17_sar} combined a number of extragalactic X-ray survey fields with multi-wavelength data to construct a non-parametric model of the specific accretion rate distribution. Their methodology required that the convolution of the P($\lambda_{\rm{sBHAR}}$) with the galaxy stellar mass function yields the observed number of X-ray AGN in bins of luminosity, redshift and stellar mass. \cite{aird18_sar} started with a sample of near-infrared selected galaxies for which stellar masses and SFRs were estimated. X-ray observations were then used to extract the X-ray photons at the positions of individual galaxies. These were then fed into a flexible Bayesian mixture model to determine in a non-parametric manner the corresponding specific accretion rate distribution of star-forming and quiescent galaxies as a function of stellar mass and redshift. Despite differences in the methodology the \cite{age17_sar} and \cite{aird18_sar} constraints on the specific accretion rate distribution are in good agreement \citep[see][]{age17_sar}. Both \cite{age17_sar} and \cite{aird18_sar} measured P$(\lambda_{\rm{sBHAR}})$ as a function of redshift out to $z\approx3$. Figure~\ref{fig:sar} graphically shows examples of the specific accretion rate distributions used in our analysis.

Using these distributions we associate AGN X-ray luminosities to galaxies in the \textsc{UniverseMachine} catalogues. This process is done in a probabilistic approach. For each mock galaxy with stellar mass $M_\star$ and redshift $z$ the corresponding specific accretion rate distributions from either \cite{age17_sar} or \cite{aird18_sar} are sampled to draw random $\lambda_{\rm{sBHAR}}$. These are then used to assign X-ray luminosities to individual galaxies by inverting Equation~(\ref{eq:sar_def}). \cite{aird18_sar} provides separate P$(\lambda_{\rm{sBHAR}})$ for star-forming and passive galaxies. In this case we split the \textsc{UniverseMachine} galaxies into these two classes using the relation of \cite{aird18_sar}
 
\begin{equation}
    \log \rm{SFR}_{\rm{cut}} = -8.9 + 0.76 \log \left(\frac{M_\star}{M_\odot}\right) + 2.95 \left( 1 + \it{z} \right).
    \label{eq:sfr_cut}
\end{equation}

\noindent At fixed stellar mass and redshift galaxies with star-formation rate above and below $\rm{SFR}_{\rm{cut}}$ are considered star-forming and passive respectively.  

The extragalactic survey fields used by \cite{age17_sar} and \cite{aird18_sar}  are dominated by low density regions of the cosmic web. Groups and cluster of galaxies, although present in these samples, are subdominant simply because of the form of the halo mass function and the relevantly small FOV of most of the survey fields used. The derived specific accretion rate distributions are therefore representative of the field galaxy population, i.e. those outside massive groups or clusters of galaxies. For this reason in what follows we refer to the predictions of the model as ``field expectation''. The adopted specific accretion rate distributions are agnostic to the parent halo mass of individual galaxies, hence the the zero-order assumption of the model that the incidence of AGN in galaxies is independent of environment.

\begin{figure*}
	\includegraphics[width=1.75\columnwidth]{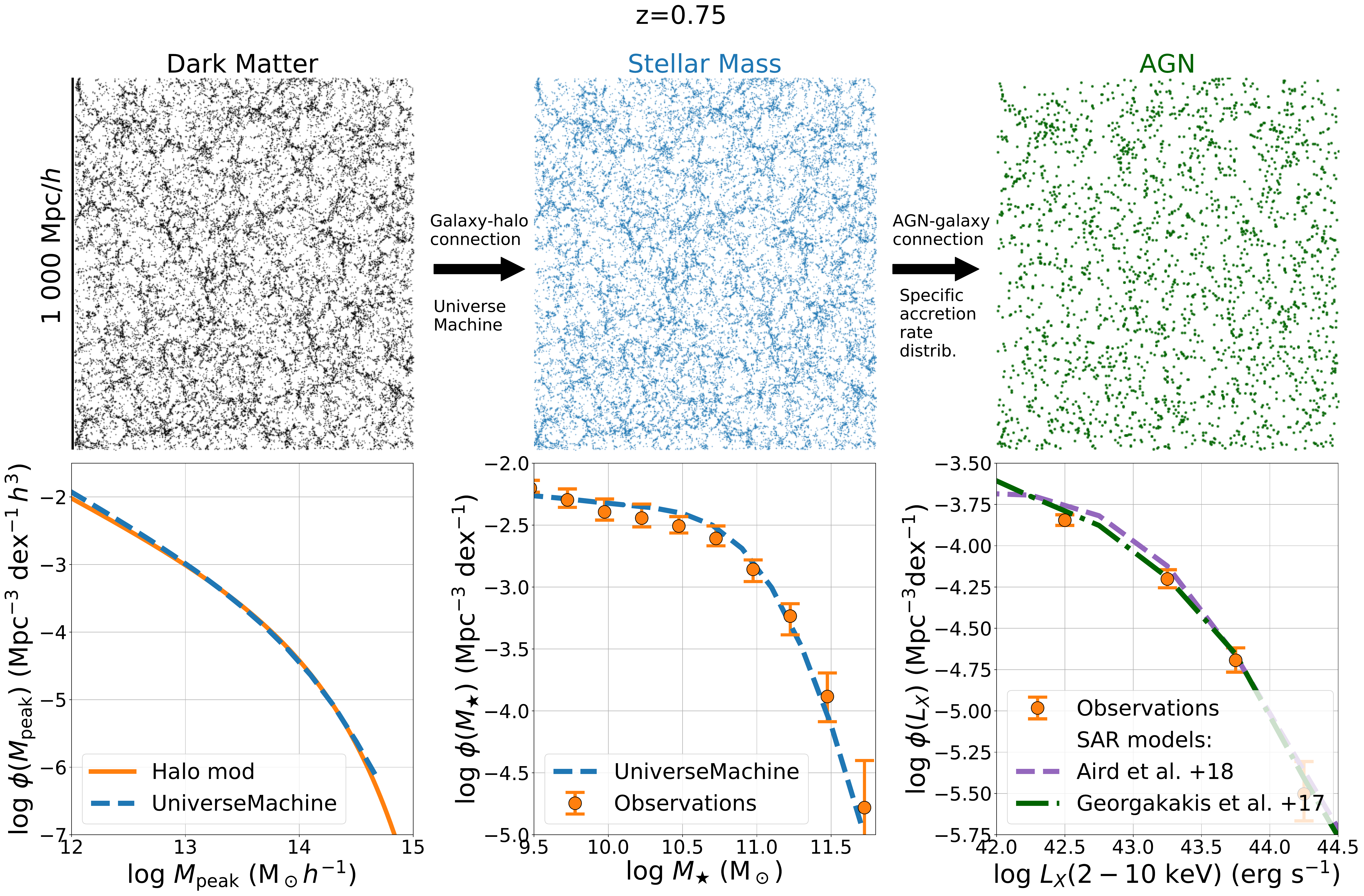}\\
    \caption{Graphical workflow of the the semi-empirical modeling to construct mock AGN catalogues based on dark matter N-body simulations. The top panels correspond to a 10 Mpc/$h$ slice of a box from MDPL2 cosmological simulation with 1\,000 Mpc/$h$ side size at the snapshot $z=0.75$. The dots represent the  positions of dark matter haloes with masses $M_{\rm peak}>10^{12}\,M_\odot/h$ (top left panel), galaxies within these dark matter haloes (top middle panel) and AGN within the same dark matter haloes (top right panel). Only AGN with $L_X\,(2-10\,\rm{keV})>10^{42}\,\rm{erg\,s^{-1}}$ using the specific accretion distribution described in \citet[]{age17_sar}. The construction of AGN mocks proceeds from left to right in this figure. Dark matter halos (black dots in the top left panel) are populated with galaxies (blue points in the top middle panel), using abundance matching \citep[\textsc{UniverseMachine}, ][]{behroozi19_um}. These galaxies are seeded with accretion events following the observationally derived distributions of these events \citep[e.g.][]{age17_sar,aird18_sar}. The key feature of this approach is the reproduction by construction of the predictions and observables shown in the lower set of panels. Lower left panel: halo mass function predicted by theoretical models in orange (HaloMod, \url{https://pypi.org/project/halomod/}) and simulations in blue \citep[MDPL2,][]{MultiDark2016}. Lower center panel: stellar mass function where circles represent observations \citep[][]{muzzinsmf2013,ilbert13_smf} and the dashed curve the semi-empirical model prediction. Lower right panel: X-ray luminosity function where circles represent observations \protect\cite[][]{age17_sar} and the curves represent the two independently derived models using the specific accretion rate distributions of \protect\citet[][ green dot-dashed]{age17_sar} and \protect\citet[][ purple dashed]{aird18_sar}}.
    \label{fig:cube}
\end{figure*}

The final products of the AGN seeding process are MDPL2 cosmological boxes at fixed redshift with galaxies (from \textsc{UniverseMachine}) and AGNs (from random sampling of the $\lambda_{\rm{sBHAR}}$ distribution). Figure~\ref{fig:cube} graphically demonstrates that our semi-empirical methodology by construction reproduces the halo mass function of dark matter haloes, the stellar mass function of galaxies and the AGN X-ray luminosity function. It is also demonstrated that this approach produces AGN mocks with the large-scale clustering ($\ga 1$\,Mpc) consistent with observations \citep{age19_mock, aird21_mock}. Moreover, the AGN duty cycle, defined as the probability of galaxies above a given stellar mass hosting an AGN above a given accretion luminosity, are inherent in the derivation of specific accretion rate distributions above. As a result our AGN and galaxy mocks are consistent with independently derived determinations of the AGN duty cycles \citep[e.g.][]{Goulding2014}. Put differently, the stellar mass function of the mock AGN host galaxies at fixed X-ray luminosity threshold is consistent with observational constraints \citep{Georgakakis2011, age17_sar}. 

It is noted that the above methodology for seeding galaxies and halos with AGN is similar to that proposed by \cite{Shankar2020} and \cite{allevato21} for generating mock AGN samples based on the semi-empirical approach. In these studies satellites and central galaxies/AGN can be treated separately by changing their relative duty cycles. In our approach there is no distinction between the two, i.e. it is assumed that both central and satellites are described by the same duty cycle.

\subsection{Light-cones}\label{sec:lc}
The comparison of the predictions from the simulations with the observations is following the principles of forward modeling. For that the \textsc{UniverseMachine} boxes need to be first projected onto the sky plane to mimic real observations. We assume a box with a XYZ Cartesian coordinate system. This is offset along the Z-axis by the co-moving distance $D_c(z)$ at redshift $z$. The observer is placed at $\rm{Z}=0$\,Mpc$/h$ and on the XY plane. The line-of-sight angle of the observer relative to every galaxy in the box is then estimated. This angle can be split into a right ascension and declination on the unit sphere. The redshift of each object corresponds to its co-moving distance from the observer. The finite FOV of real observations can also be imposed by defining a sight-line from the observer to a given light-cone direction, estimating the angular distances of all mock galaxies relative to this direction and then rejecting the ones with angular distances larger than the adopted FOV.

For the analysis presented in this paper we construct light-cones in the vicinity of clusters. We consider three redshifts $z=0.2, 0.75$ and 1.25, which correspond to the mean redshifts of the cluster samples presented by \cite{martini09, martini13}. We select \textsc{UniverseMachine} boxes with scale factor\footnote{Defined as $a=1/(1+z)$.} 0.4505, 0.5747 and 0.8376 that approximately correspond to each of the redshifts above. For a given box a massive dark matter halo is selected (see Section \ref{sec:selec_eff} for details) and is placed at a co-moving distance $D_c(z_c)$ from the observer, where $z_c$ is the box redshift, i.e. one of 0.2, 0.75 and 1.25. The light-cone to the cluster is then constructed with an opening angle that is defined by the user (see next section for details). The end product are dark matter haloes, mock galaxies and AGN projected on the sky that mimic real observations.

\subsection{Selection effects}\label{sec:selec_eff}
This section describes how observational selection effects are implemented into our simulations to allow comparison with the results of \cite{martini09,martini13} on the fraction of AGN in galaxy clusters. The characteristics of the observations we attempt to mimic can be grouped into three broad categories that relate to the richness/mass of the cluster sample, the galaxy/AGN cluster membership criteria and the apparent brightness or stellar mass of the galaxy/AGN sample. Below we discuss each of them in detail. 

\subsubsection{Cluster sample}\label{sec:clust_sample}
We define the cluster sample by adopting a minimum virial mass threshold.  \cite{martini09} provide velocity dispersion for their cluster sample as a measure of their masses. However, the \textsc{UniverseMachine} dataset does not include velocity dispersion information and therefore a mapping is required between this parameter and halo mass. For the latter we adopt the analytical relations presented by \cite{munari13_disp_hm_rela} based on N-body simulations. This allows us to associate individual \textsc{UniverseMachine} halos with a velocity dispersion and then threshold on this  quantity to mimic the \cite{martini09} cluster sample selection. We choose the lower halo mass limit in such a way that our parent cluster sample reproduces the median velocity dispersion of the \cite{martini09} sample. The adopted virial halo mass limits are $5.7\cdot10^{14}$ M$_\odot/h$ and $3.6\cdot10^{14}$ M$_\odot/h$ for $z=0.2$ and 0.75 respectively. For the high-redshift clusters of \cite{martini13} there is only scattered information on their halo masses. Literature results suggest masses of few times $10^{14}\,M_\odot$, based on dynamical measurements or estimates from X-ray luminosities. For our high-redshift simulations ($z=1.25$) we therefore select halos with virial mass $>3\cdot10^{14}\,M_\odot/\it{h}$ to mimic the cluster sample of \cite{martini13}. Our results and conclusions are not sensitive to this threshold.

At the mass limits above there are 388, 157 and 18 parent halos in the MDLP2/\textsc{UniverseMachine} boxes at redshifts 0.2, 0.75 and 1.25 respectively. These numbers exclude halos close to the box edges, whose volume as defined in observations (see text below for more details), intersects the box boundaries. The rapid decrease in the number of clusters in each sample is because of the strong evolution of the halo mass function with redshift. These clusters are then projected onto the sky as described in Section \ref{sec:lc}. We choose to place the observer at the same (X,Y) position as the cluster  with respect to the reference system of the box. This results in light-cones centered on the each of the selected massive halos, with a line-of-sight perpendicular to the (X,Y)-plane of the box. We define the FOV in terms of the virial radius of the cluster.

\subsubsection{Cluster membership}
Although in the simulation box the satellite galaxies associated with a given parent halo are known, we prefer to follow an observational-motivated approach for defining cluster membership based on the projected and radial distances relative to the cluster center. Cluster member candidates are selected to lie within the projected $R_{\rm vir}$ radius of the corresponding halo. This is similar to the selection of \cite{martini09,martini13}. We choose to use $R_{\rm vir}$ instead of $R_{200}$ since they are similar, but the last is not present in \textsc{UniverseMachine} dataset. We checked that this assumption does not affect our final results and conclusions.   

\begin{figure}
    \centering 
    \includegraphics[width=.49\textwidth]{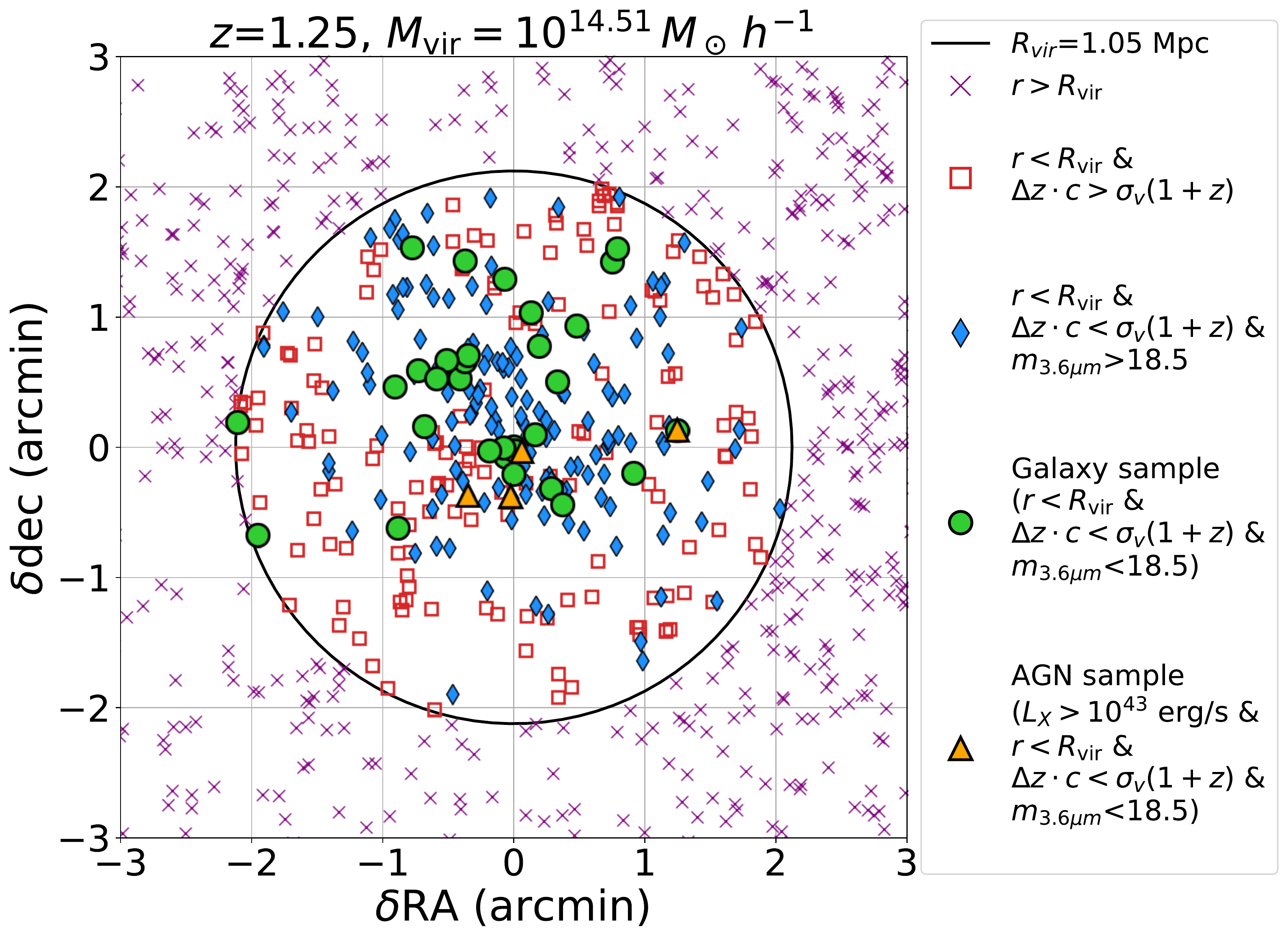}
    \caption{The light-cone of a massive cluster in \textsc{UniverseMachine} at redshift $z=1.25$ and halo mass $M_{\rm{vir}}=10^{14.51}M_\odot/h$. Large circle marks the virial radius of the cluster, all symbols correspond to \textsc{UniverseMachine} galaxies within the light-cone. Different symbols and colors demonstrate selection effects as described in Section~\ref{sec:selec_eff}. Purple crosses, red empty squares and blue diamonds indicate galaxies that are excluded from the sample because of the selection effects as indicated in the legend. Green circles and orange triangles indicate the final sample of galaxies and AGNs with $L_X(2-10\,\rm keV)>10^{43}$ erg/s after applying all the selection effects.}
    \label{fig:fov_cuts}
\end{figure}

The radial distance of clusters member candidates relative to the cluster centre is measured in redshift space as in real observations. Cluster members are those mock galaxies or AGN with redshift difference to the cluster $\Delta z\cdot c \leq 3\sigma_v(1 + z)$, where $z$ is the redshift of the cluster (fixed to be one of the redshifts of interest, i.e. $z=0.2$, 0.75 or 1.25), $c$ represents the speed of light and $\sigma_v$ is the velocity dispersion of the cluster determined using \cite{munari13_disp_hm_rela}. This definition corresponds to the selection criteria adopted by \cite{martini09} for defining cluster members and it is used for the clusters at redshifts $z=0.2$ and 0.75. In the case of \cite{martini13} 3$\sigma_v$ is fixed to $2\,000\, {\rm km/s}$ for all the clusters. This restriction is also adopted in our simulation for the clusters at redshift $z=1.25$. This condition defines the volume of the cluster in terms of its velocity dispersion. Figure~\ref{fig:fov_cuts} shows an example of a simulated observation and demonstrates the impact of selection effects.

\subsubsection{Galaxy/AGN sample selection}\label{sec:sample_select}

\begin{table}
    \centering
    \caption{Selection effects adopted for defining the mock cluster galaxy and AGN samples.}
    \label{tab:sel_effects_values}
    \begin{tabular}{c|c|c|c|c|c}
         $z_{\rm box}$ & $M_{\rm vir,lim}$ & $m_{\rm lim}$  & band & $M_{\star, \rm lim}$ & L$_{\rm X, lim}$ \\
         (adim.)& ($M_\odot/h$) & (mag)  & (adim.) & ($M_\odot$) & (erg\,s$^{-1}$)\\
         (1) & (2) & (3) & (4) & (5) & (6) \\\hline 
         1.25 & $3\cdot10^{14}$ & 18.5  & IRAC1 & 2.2$\cdot$10$^{10}$ & $10^{43}$ and $10^{44}$ \\
         0.75 & 3.6$\cdot10^{14}$ & 23.3 & R band & 2.2$\cdot$10$^{10}$ & $10^{43}$ and $10^{44}$\\
         0.2  & 5.7$\cdot10^{14}$ & 19.1 & R band & 2.2$\cdot$10$^{10}$ & $10^{43}$ and $10^{44}$\\ \hline \hline
    \end{tabular}
    \begin{list}{}{}
    \item  (1) Redshift of each cluster corresponding to those of \citet[][]{martini09,martini13}, (2) Minimum dark matter halo mass (virial) adopted to define the parent cluster sample at different redshifts, (3) minimum magnitude threshold adopted to define galaxy/AGN cluster members at different redshifts, (4) Photometric band used to define the magnitude threshold, (5) Stellar mass limit used to select galaxy/AGN cluster members at different redshifts and (6) X-ray luminosity limit adopted to define the AGN sample at different redshifts.
    \end{list}
\end{table}

In observations AGN and/or galaxy samples are typically selected above a given apparent magnitude limit. In \cite{martini09,martini13} for example, this is set relative to the knee of the optical and/or mid-infrared luminosity function of galaxies at the corresponding cluster redshift (see Section~\ref{sec:obs}). In simulations however, like the semi-empirical model described in this work, galaxies are defined by their intrinsic properties, such as stellar mass, SFR and accretion luminosity. Associating these physical properties to apparent magnitudes requires assumptions on e.g. the SFH of galaxies, the spectral energy distribution (SED) of stellar populations or the shape and normalization of the dust attenuation curve. Assigning SEDs to simulated galaxies is therefore far from trivial and inevitably requires additional modelling steps \citep[e.g.][]{ueda20,pearl21_remapSFR}

Our baseline model/observation comparison avoids these additional steps. Instead we make the simplifying assumption that the knee of the observed galaxy optical or mid-infrared luminosity function traces the knee of the underlying galaxy stellar mass function, $M^*_\star$. This allows us to translate the $R$-band and $3.6\,\rm \mu$m apparent magnitude limits of $0.4\,L^*$ adopted by \citet[see Section \ref{sec:obs}]{martini09,martini13} to stellar mass cuts. Mock galaxies are selected to be more massive than $\log\,M^*_\star/M_\odot - 0.4\,\rm dex$. The break of the mass function  is fixed to  $\log\,M^*_\star=10^{10.7}\,M_\odot$ independent of redshift based on the parametrisation of \cite{ilbert13_smf}. Although the \cite{ilbert13_smf} study refers to field galaxies, observations show that the shape of stellar mass function is similar in massive clusters  \citep[e.g.][]{Vulcani2013, Nantais2016}.  The translation of the $0.4\,L^*$ apparent magnitude limit to a $\log\,M^*_\star/M_\odot-0.4\,\rm dex$ threshold implies the same average mass-to-light ratio for galaxies. This approximation is justifiable in the case of the high redshift cluster sample ($z\approx1.25$) of \cite{martini13}, where galaxies are selected in the $IRAC$ $3.6\,\mu$m band. At the mean cluster redshift, this wavelength roughly corresponds to rest-fame near-infrared ($\approx$ 1.6$\,\mu$m), where the mass-to-light ratio is not a strong function of the galaxy stellar population. We acknowledge that in the low-redshift ($z=0.2$ and 0.75) cluster sample of \cite{martini09}, galaxies are selected in the $R$-band, where variations of the mass-to-light ratio as a function of the star-formation rate are important. For this sample the approximation of a constant mass-to-light ratio is rough and should be taken with caution.
  
We further address the limitations above by assigning apparent magnitudes to mock galaxies in the light-cones following the methodology described in \cite{ueda20}. \textsc{UniverseMachine} galaxies are assumed to be described by exponentially declining SFH. The parameters of the SFH model are constrained to reproduce the \textsc{UniverseMachine} stellar masses and instantaneous SFRs of the galaxies at their assigned redshifts in the light-cone. The \cite{BruzualCharlot03} stellar library and the \cite{chabrier03} initial mass function are used to synthesize stellar populations for the adopted SFH. The SEDs of star-forming galaxies are extincted by dust assuming the \cite{calcetti2000} law and $E(B - V)$ = 0.4 mag. The magnitudes of passive galaxies are not extincted by dust. This empirical model is shown to reproduce reasonably well the distribution of apparent magnitudes of galaxies in the COSMOS field \citep{muzzinsmf2013}.

The assigned apparent magnitudes are sensitive to the instantaneous SFR of mock galaxies. The empirical model of \cite{ueda20} assumes that star-forming galaxies are on the Main Sequence of star-formation \citep[][]{schreiber+15}, while quiescent galaxies lie $1-1.5$\,dex below it depending on redshift. This offset for the quenched galaxies is empirically determined to reproduce the observed magnitude distribution of passive galaxies as a function of redshift in the COSMOS field. The \textsc{UniverseMachine} star-forming galaxies are also constrained to follow the Main Sequence of star-formation at different redshifts and are therefore consistent with the assumptions of the empirical SED model of \cite{ueda20}. In contrast, the SFR distribution of quenched galaxies is not as well constrained by observations \citep[see discussion by][]{pearl21_remapSFR}. Passive galaxies in \textsc{UniverseMachine} are assigned specific star-formation rates (sSFR) that are drawn from a non-evolving log-normal distribution with mean $\log \rm{sSFR}/\rm{yr}=-11.8$ and scatter 0.36\,dex, motivated by observations in the local Universe \citep[][]{behroozi15}. This SFR is 2\,dex below the main sequence of star-formation at low redshift and therefore inconsistent with the assumptions of the empirical SED model of \cite{ueda20}. This difference in sSFR results in passive galaxies with too faint apparent magnitudes for the empirical model of \cite{ueda20}. In this work we therefore adopt the definition of quenched galaxies given by Equation~\ref{eq:sfr_cut} 
but then re-scale the corresponding sSFR according to the \cite{ueda20} prescriptions (i.e. 1$-$1.5\,dex below the main should that be $M_R^*(z=0)$ sequence) before assigning them magnitudes.

Once apparent magnitudes are assigned to mock galaxies in the light-cone, we apply cuts similar to those adopted by \cite{martini09,martini13}, i.e. one magnitude fainter than the break of the luminosity function in the $R$ (low- and medium-redshift cluster samples, $z=0.2$ and 0.75) and 3.6$\,\mu$m (high-redshift cluster samples; $z=1.25$) bands. The magnitude limits for the different samples are summarized in Table~\ref{tab:sel_effects_values}. For the $R$-band in particular, following \cite{martini09} we assume that the knee of the luminosity function evolves with redshift as $M_R(z)=M_R(z=0)+z$ with $M_R^*(z=0)=-21.92$ \citep[][]{christlein03_break_lf}. This absolute magnitude is converted to apparent magnitude in the $R$-band assuming an elliptical galaxy SED \citep[][]{ilbert09_photoRed} for the K-corrections. The estimated knee of the luminosity function in apparent magnitudes is  $m_R^*=17.1$, 22.3\,mag  at $z= 0.2$ and 0.75 respectively. The limits listed in Table \ref{tab:sel_effects_values} are one magnitude fainter than the apparent magnitude of knee of the optical luminosity function at the relevant redshift. For the 3.6\,$\mu$m band the knee of  the luminosity function is assumed to 17.5 mag (apparent magnitude) at $z=1.25$ \citep[][]{mancone10}. Mock galaxies and AGNs in the simulated light-cones are selected to be brighter than the magnitudes limits listed in Table \ref{tab:sel_effects_values}. For the mock AGN we further apply the X-ray luminosity limits of \cite{martini09,martini13} to define two samples with $L_X(\rm 2-10\,keV) > 10^{43}\,\rm{erg \,s^{-1}}$ and $\rm >10^{44}\,\rm{erg \,s^{-1}}$.

In the analysis above we ignore the contribution of AGN light to the observed magnitude of galaxies. Modeling this component requires knowledge of the obscuration properties of the AGN. This latter parameter has the strongest impact on the observed spectral energy distribution of these sources. The specific accretion rate distribution used in this work to seed galaxies with AGN luminosities do not account for the AGN obscuration and therefore cannot be used to model this effect \citep[see also][]{ueda20}. We simplify this problem by ignoring the AGN contribution to the emission of the galaxy in the optical/mid-infrared. This simplification is equivalent to the assumption that mock AGN are completely obscured and hence subdominant relative to the stellar emission of galaxies. We will discuss the impact of this assumption on the results and conclusions in the next sections.

\section{Results}\label{sec:results}
\begin{figure*}
    \centering 
    \includegraphics[width=1\textwidth]{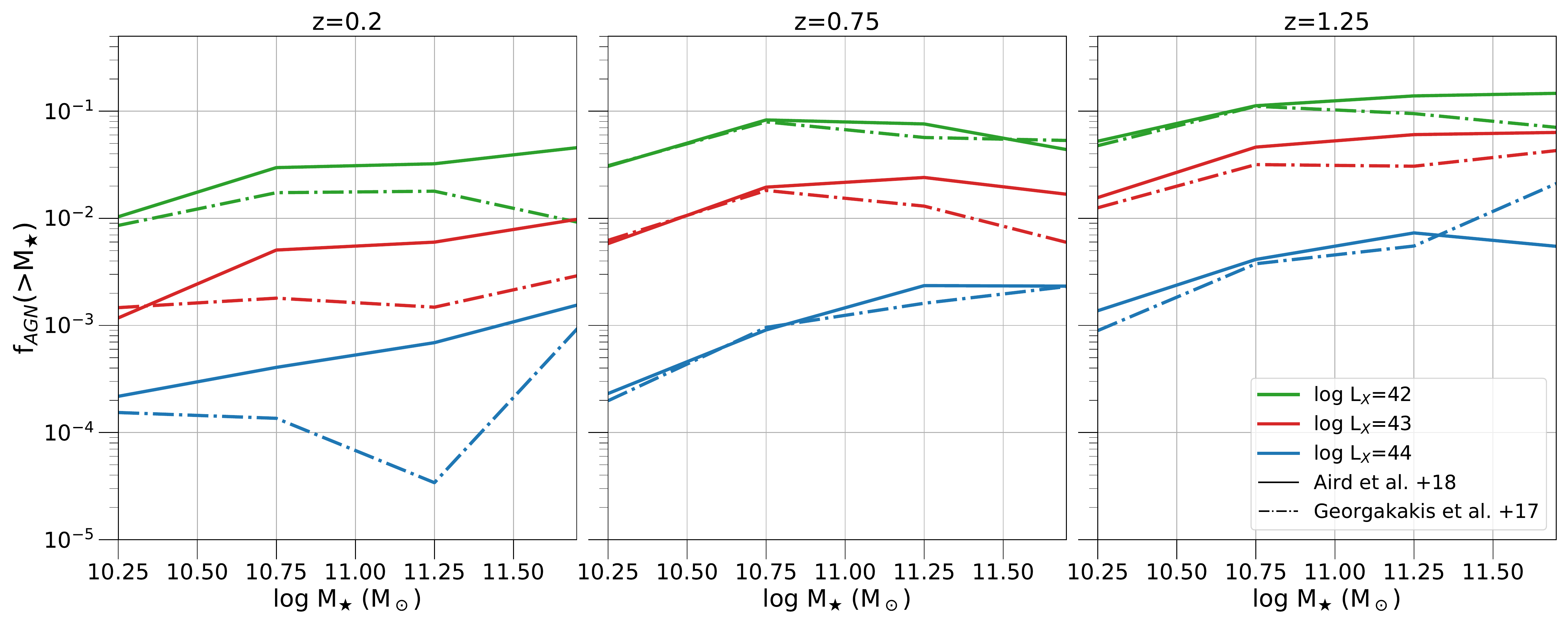}
    \caption{Fraction of AGN relative to galaxies above a stellar mass limit as a function of stellar mass. The curves correspond to the predictions of the semi-empirical model described in Section \ref{sec:agn_galax_conec}. Different colours correspond to AGN more luminous than $L_X(\rm 2-10\,keV)>10^{42}\,erg\,s^{-1}$ (green), $\rm >10^{43}\,erg\,s^{-1}$ (red) and $\rm >10^{44}\,erg\,s^{-1}$ (blue). Different line styles indicate different specific accretion rate distribution models adopted to generate the AGN mocks (see Section~\ref{sec:agn_galax_conec}). The dashed-dotted lines are for the \citet{age17_sar} and the solid lines correspond to the \citet{aird18_sar} specific accretion-rate distributions. Each panels correspond to redshifts from left to right of $z=0.2$, 0.75 and 1.25.}
    \label{fig:frac_SM}
\end{figure*} 
\subsection{Impact of selection effects on the incidence of AGN}\label{sec:results_impact_select_eff}
We first explore the sensitivity of the estimate AGN fractions to observational selection effects. Figure  \ref{fig:frac_SM} plots the AGN duty cycle of our semi-empirical model as a function of host galaxy stellar mass limit. The former quantity is defined as the fraction of mock AGN above a given X-ray luminosity threshold among galaxies more massive than a given stellar mass limit (x-axis of Figure \ref{fig:frac_SM}). In this exercise all galaxies in a given \textsc{UniverseMachine} box are used independent of halo mass. We iterate that the AGN fractions plotted in Figure \ref{fig:frac_SM}  are consistent with independent observations that use the stellar mass function of galaxies and AGN hosts to infer duty cycles \citep[see][]{age17_sar}. Figure \ref{fig:frac_SM} shows that the AGN duty cycle is sensitive to both the X-ray and the host galaxy stellar mass thresholds. There is a  general trend of increasing AGN fraction with decreasing X-ray luminosity. This is because less luminous AGN are more common than high accretion luminosity events (i.e. the X-ray luminosity function). Also the AGN duty cycle at fixed luminosity drops with decreasing stellar mass below about $10^{11}\,M_{\odot}$. This is because in our implementation lower stellar mass hosts require higher $\lambda_{\rm{sBHAR}}$ to produce AGN above a given luminosity cut. However, higher specific accretion rates are less likely. This is demonstrated by the form of the specific accretion rate distribution plotted in Figure \ref{fig:sar}, which strongly decreases with increasing $\lambda_{\rm{sBHAR}}$. For stellar masses $\ga 10^{11}\,M_{\odot}$ the duty cycle curves of Figure~\ref{fig:frac_SM} either increase, decrease or remain nearly flat with increasing stellar mass. This is related to the stellar mass dependence of the specific accretion-rate distributions of \cite{age17_sar} and \cite{aird18_sar} used in our analysis. It is also worth noting that the differences between the two specific accretion rate distributions models above are stronger for the lowest redshift panel ($z=0.2$). This is related to the small sample of low redshift AGN in the \cite{age17_sar} and \cite{aird18_sar} studies. In any case, the important point of  Figure~\ref{fig:frac_SM} is that AGN fractions are sensitive to the choice of the stellar mass and X-ray luminosity thresholds, i.e. the observational selections effects. This emphasises the importance of carefully treating this issue, as described in Sections \ref{sec:sample_select}.

\begin{figure*}
    \centering 
    \includegraphics[width=1\textwidth]{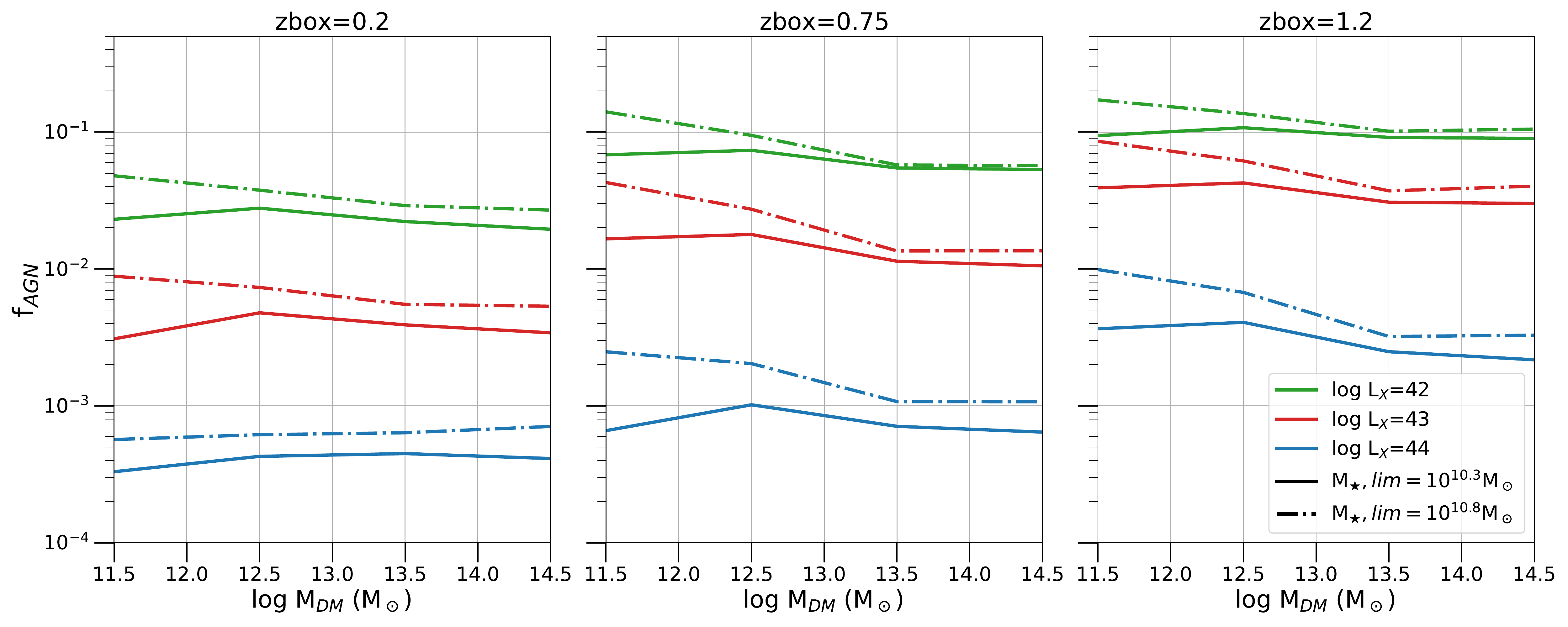}
    \caption{Fraction of AGN in galaxies as a function of parent halo mass. The curves correspond to the predictions of the semi-empirical model described in Section \ref{sec:agn_galax_conec} using the specific accretion rate distribution from \citet{aird18_sar}. Different colours correspond to AGN more luminous than $L_X(\rm 2-10\,keV)>10^{42}\,erg\,s^{-1}$ (green), $\rm >10^{43}\,erg\,s^{-1}$ (red) and $\rm >10^{44}\,erg\,s^{-1}$ (blue). Different line styles indicate different stellar mass limit cuts M$_{\star,lim}$>10$^{10.3}$ M$_\odot$ (solid line) and M$_{\star,lim}$>10$^{10.8}$ M$_\odot$ (dashed-dotted line). Each panel corresponds to redshifts from left to right of $z=0.2$, 0.75 and 1.25.}
    \label{fig:frac_HM}
\end{figure*}

For completeness we also show in Figure \ref{fig:frac_HM} how the fraction of AGN varies with parent halo mass. \textsc{UniverseMachine} galaxies (central or satellites) above a given stellar mass threshold are grouped by parent halo mass. At a fixed halo mass the fraction of galaxies that host AGN above a given accretion luminosity limit is estimated and plotted. As expected the resulting curves are nearly flat with halo mass since the explicit assumption of the semi-empirical model construction is that the probability of a galaxy hosting an accretion event is agnostic to halo mass. Nevertheless, the strong correlation between halo and stellar mass as well as the X-ray luminosity cut imprint systematic trends onto the curves of  Figure~\ref{fig:frac_HM}. These are manifested for example, by the increasing AGN fraction toward lower halo masses. This is more pronounced  for the curves with a high stellar mass cut, $\log M_{\star}/M_{\odot} > 10.8$. This is because this threshold essentially removes a large number of lower mass galaxies, which are typically found in low mass halos. Additionally the form of the specific accretion rate distributions dictates that more massive galaxies are more likely to host AGN above a fixed accretion luminosity threshold. The net effect is the observed increase in the AGN fraction toward the low halo mass end  in the case of the higher stellar mass threshold in Figure~\ref{fig:frac_HM}. 

\subsection{X-ray AGN fractions in clusters}
\begin{figure*}
    \centering 
    \includegraphics[width=.8\textwidth]{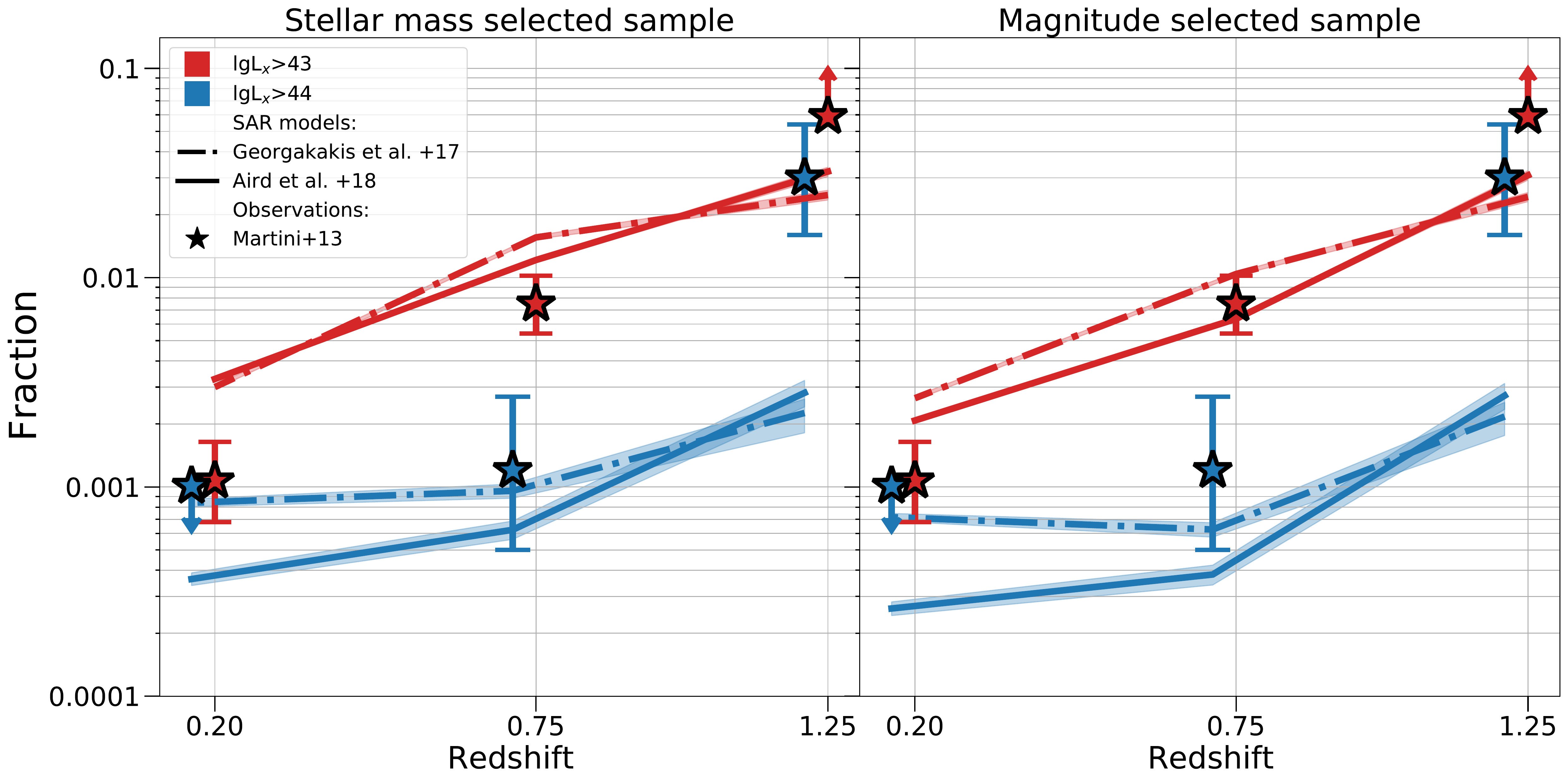}
    \caption{Evolution of the X-ray AGN fractions in massive clusters. The observations (stars) and semi-empirical model predictions (lines and shaded regions) are plotted at redshifts $z = 0.2, 0.75$ and 1.25. In both panels the stars (red or blue) are the observationally measured AGN fractions of \citet[][see Section \protect\ref{sec:obs}]{martini13}.  Different colours indicate different luminosity cuts. Red is for AGN with $L_X (\rm 2-10\, keV)> 10^{43}\,erg\,s^{-1}$ and blue corresponds to $L_{X} \rm (2-10\, keV)> 10^{44}\,erg\,s^{-1}$. For the shake of clarity the blue stars are shifted by -0.12 in the x-axis direction. The errorbars correspond to the 1$\sigma$  error. The lower limit at $z=1.25$ is due to incompleteness and the upper limit at $z=0.2$ corresponds to the 3$\sigma$ confidence level. In both panels the lines and corresponding shaded regions represent the predictions of the semi-empirical model on the AGN fraction under different assumptions on the selection effects. On the left panel the simulated galaxy/AGN samples are selected above the stellar mass limit listed in Table \ref{tab:sel_effects_values}. The right panel corresponds to mock AGN/galaxy samples selected above the apparent magnitude thresholds listed in Table \ref{tab:sel_effects_values} (see Section \ref{sec:selec_eff} for more details). Lines of different colour indicate different luminosity cuts as explained above. The blue lines should therefore be compared with the blue stars and the same for the red lines/symbols. Different line styles indicate different specific accretion rate distribution models adopted to generate the AGN mocks (see Section~\ref{sec:agn_galax_conec}). The dashed-dotted lines are for the \citet{age17_sar} and the solid lines correspond to the \citet{aird18_sar} specific accretion-rate distributions. The shaded regions within which lines are embedded correspond to the 68\% confidence intervals of the mean value calculated using the bootstrapping technique described in the text.}
    \label{fig:fractions}
\end{figure*}

Having demonstrated the strong impact of selection effects on the calculation of AGN fractions, we next turn to the comparison of the our model predictions with the observed fractions of AGN in massive galaxy clusters. Figure~\ref{fig:fractions} plots the fraction of AGN among cluster member galaxies as a function of redshift. The observational results of \cite{martini09,martini13} at mean redshifts $z = 0.2$, 0.75 and 1.25 are compared with the predictions of our semi-empirical model using either the \cite{aird18_sar} or the \cite{age17_sar} specific accretion rate distributions. The model predictions for the mass- and magnitude-limited samples are presented in different panels. Cluster AGN fractions are estimated for two luminosity thresholds, $L_X(\rm 2-10\,keV)>10^{43}\,\rm{erg\,s^{-1}}$ and $>10^{44}\,\rm{erg\,s^{-1}}$ indicated by different colors.  The uncertainties assigned to the model fractions are determined using bootstrap resampling. For each cluster a total of 10 AGN realisations are generated by repeating the seeding process of Section \ref{sec:agn_galax_conec} 10 times (re-seeded samples). For a given cluster these 10 realisations differ in their AGN populations because of the stochastic nature of the seeding process. This results in an extended cluster sample that is 10 times larger than the original, i.e. 3880, 1570, 180 for $z=0.2$, 0.75 and 1.25 respectively. At fixed redshift clusters are drawn with replacement from the extended sample to generate a total of 100 sub-samples which are used to determine the 68\% confidence interval around the mean. These confidence intervals are the errobars of the model predictions plotted in Figure~\ref{fig:fractions}. They represent the uncertainty of the mean expected fraction of AGN per cluster. 

The different model flavors broadly yield consistent results at fixed redshift and X-ray luminosity threshold. At the lowest redshift bin however, differences are apparent between the AGN fractions at $L_X(\rm 2-10\,keV)>10^{44}\,erg\,s^{-1}$  for the models using the \cite{aird18_sar} and \cite{age17_sar} specific accretion rate distributions. 

These differences are ultimately linked to the observationally measured specific accretion rate distributions and their dependence on the stellar mass of the AGN hosts. The probability of high mass galaxies hosting an accretion event is lower in the \cite{aird18_sar} specific accretion rate distributions compared to the \cite{age17_sar} ones. As a result the AGN fractions predicted by the model flavor that uses the \cite{age17_sar} specific accretion rate distributions are higher. This is primarily because of luminous AGN assigned to the central galaxies of the clusters. About 42 out 388 ($\sim$11\%) of the mock clusters have a central galaxy with assigned AGN luminosity $L_X(\rm 2-10\,keV)>10^{44}\,\rm{erg\,s^{-1}}$. Such a high incidence AGN is inconsistent with observational constraints on the X-ray properties of Brightest Cluster Galaxies in local massive clusters \citep[][]{yiang18_bcgs}. This is a limitation of the  \cite{age17_sar} observationally determined specific accretion rate distributions. 

The fractions predicted by the semi-empirical model flavor with the apparent magnitude selection effects (right panel in Figure~\ref{fig:fractions}) does not include the contribution of AGN emission to the mock galaxy SED. In the case of unobscured AGN this contribution may dominate over the host galaxy stellar component in the optical/mid-infrared bands, particularly at high accretion luminosities ($L_X \rm \ga10^{44}\, erg \, s^{-1}$).  The estimated model fractions may therefore be underestimated, because a higher fraction of mock AGN would be brighter than the adopted apparent magnitude cut, if their contribution to the mock galaxy SED was modeled. We estimate nevertheless that this effect is small. We modify the methodology of Section~\ref{sec:selec_eff} by including all AGN cluster members above the luminosity limits $L_X(\rm 2-10\,keV)>10^{43}\,\rm{erg\,s^{-1}}$ or $>10^{44}\,\rm{erg\,s^{-1}}$ in the calculation of the corresponding AGN cluster fractions, irrespective of the apparent magnitude of their host galaxy. This is equivalent to assuming that all of the mock AGN are unobscured and therefore their apparent magnitudes dominate that of their hosts. This approach nevertheless increases the estimated fraction at any redshift by $\la 0.1$\,dex and therefore does not impact our results and conclusions. 

    \begin{table}
        \caption{Fraction of X-ray AGN relative to galaxies in simulated massive clusters for different redshifts.}
		\centering
		\begin{tabular}{c|c|c|c|c|c}
        \multirow{2}{*}{z} & \multirow{2}{*}{L$_{\rm X,lim}$} & \multicolumn{2}{c}{f$_{\rm G+17}\cdot 10^{-2}$} & \multicolumn{2}{c}{f$_{\rm A+18}\cdot 10^{-2}$} \\\cline{3-6}
          &                 & SM & Mag & SM & Mag \\
         (1) & (2) & \multicolumn{2}{c}{(3)} & \multicolumn{2}{c}{(4)} \\\hline
         \multirow{2}{*}{1.25} & 10$^{43}$ & 2.48$^{+0.10}_{-0.13}$ & 2.43$^{+0.01}_{-0.12}$ & 3.21$^{+0.02}_{-0.17}$ & 3.08$^{+0.02}_{-0.16}$ \\ 

                              & 10$^{44}$ & 0.23 $^{+0.04}_{-0.04}$ & 0.22$^{+0.04}_{-0.04}$ & 0.28$^{+0.04}_{-0.04}$ & 0.27 $^{+0.04}_{-0.04}$ \\ \hline

        \multirow{2}{*}{0.75} & 10$^{43}$& 1.56 $^{+0.03}_{-0.03}$ & 1.04$^{+0.020}_{-0.021}$ & 1.21$^{+0.03}_{-0.03}$ & 0.635$^{+0.020}_{-0.016}$ \\

                             & 10$^{44}$& 0.096$^{+0.007}_{-0.007}$ & 0.063$^{+0.005}_{-0.005}$ & 0.062$^{+0.006}_{-0.006}$ & 0.038$^{+0.004}_{-0.004}$ \\ \hline

        \multirow{2}{*}{0.2} & 10$^{43}$ & 0.300$^{+0.007}_{-0.007}$ & 0.266$^{+0.006}_{-0.006}$ & 0.328$^{+0.008}_{-0.008}$ & 0.208$^{+0.006}_{-0.006}$ \\

                            & 10$^{44}$ & 0.084$^{+0.004}_{-0.004}$ & 0.072$^{+0.003}_{-0.003}$ & 0.036$^{+0.003}_{-0.003}$ & 0.0263$^{+0.0020}_{-0.0020}$ \\ \hline
		\end{tabular}
		\begin{list}{}{}
         \item  (1) Redshift, (2) X-ray luminosity threshold on the $2-10$ keV band, adopted for the AGN sample, in units of erg/s, (3) fractions corresponding to the \cite{age17_sar} model, for the stellar mass (SM) and magnitude (Mag) selected samples, (4) fractions corresponding to \cite{aird18_sar} model, for the stellar mass (SM) and magnitude (Mag) selected samples.
        \end{list}
    \label{tab:fractions}
    \end{table}
    
\section{Discussion}\label{sec:discussion}
In this work we study the fraction of AGN in massive clusters with a semi-empirical modelling technique that allows the generation of realistic AGN mock catalogues. Dark matter haloes from cosmological simulations are seeded with galaxies using abundance matching techniques \citep[][]{behroozi19_um}. On top of these galaxies accretion events by supermassive black holes are painted. The latter step is based on state-of-the-art specific accretion rate distributions derived from observations \citep[][]{age17_sar,aird18_sar}. The zero-order assumption of the semi-empirical model is that the incidence of accretion events in galaxies is independent of environment, i.e. galaxies are seeded with AGN using the same empirical relations independent of the mass of the parent halo. We refer to the predictions of the model as  "field expectation" for the reasons explained in Section~\ref{sec:agn_galax_conec}. This methodology reproduces by construction the halo mass function, stellar mass function and the X-ray luminosity function as demonstrated in Figure~\ref{fig:cube}.

\begin{figure*}
    \centering 
    \includegraphics[width=1\textwidth]{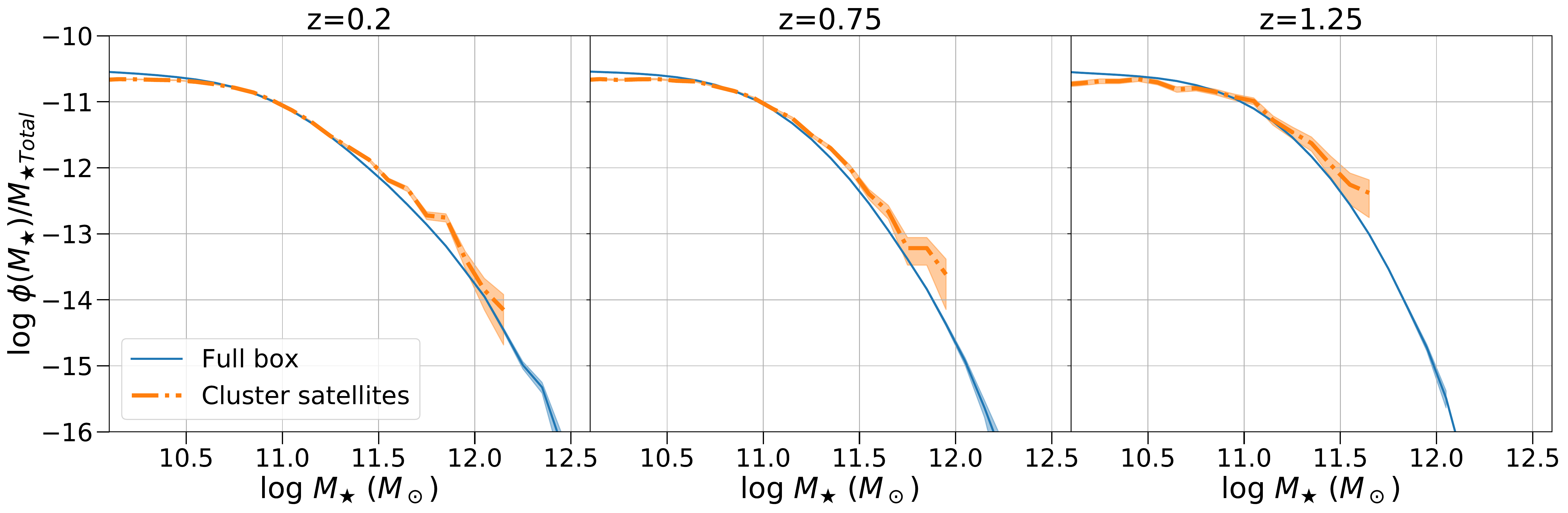}
    \caption{Stellar mass function normalized to the total stellar mass of the sample. Each panel correspond to one of the redshifts $z=0.2, 0.75$ and 1.25. The blue curve is for all galaxies in the corresponding \textsc{UniverseMachine} box and therefore represents the field stellar mass function. The orange color is the stellar mass function of the satellites of the selected clusters (see Section~\ref{sec:selec_eff} for more details). The shaded orange region correspond to the Poisson noise uncertainty.}
    \label{fig:smf_field_clust}
\end{figure*}

\subsection{AGN fractions in high redshift clusters}
A striking result in Figure~\ref{fig:fractions} is the higher observed fraction of AGN in massive clusters at $z=1.25$ compared to the model predictions. The largest discrepancy of nearly 1\,dex is for powerful AGN with $L_X(\rm 2-10\,keV) > 10^{44} \, erg \, s^{-1}$.  An enhanced fraction is also observed for moderate luminosity AGN,  $L_X(\rm 2-10\,keV) > 10^{43} \, erg \, s^{-1}$, but the fact that the observations can only place a lower limit does not allow firm conclusions on the amplitude of the effect. These findings can be attributed to systematic differences between field and cluster either in the AGN specific accretion rate distributions or the stellar mass function of the galaxy population. The former option could be interpreted as evidence for environmental dependence of the AGN triggering efficiency. The latter would indicate differences in the galaxy populations as a function of position on the cosmic web.  

Observations indicate that the shape of the total (i.e. independent of galaxy type or SFR) stellar mass function of galaxies does not strongly depend on environment out to $z\approx1.5$ \citep[e.g.][]{Vulcani2013, Nantais2016}. This behaviour is also reproduced in the \textsc{UniverseMachine} semi-empirical model. This is demonstrated in Figure \ref{fig:smf_field_clust} that compares the (average) stellar mass function of mock galaxies in the same massive clusters used in our analysis with that of all galaxies in the \textsc{UniverseMachine} box. The mass functions in Figure \ref{fig:smf_field_clust} are normalised so that their integral yields the same stellar mass density. This allows direct comparison of the mass function shapes, which are remarkably similar between massive clusters and the full box. The latter is dominated by field galaxies (i.e. not associated with massive halos) and by construction reproduces the observed stellar mass function at different redshifts estimated using extragalactic survey fields \citep[e.g.][]{muzzinsmf2013,ilbert13_smf,Moustakas_2013}. The construction of AGN mocks using the seeding process described in section \ref{sec:agn_galax_conec} is primarily sensitive to the shape of the galaxy mass function. The evidence above therefore suggests that the difference between observations and semi-empirical model predictions in Figure \ref{fig:fractions} cannot be attributed to systematic variations of the total stellar mass function with environment.

We acknowledge differences between field and cluster mass functions for star-forming and quiescent galaxies \citep[e.g.][]{Vulcani2013, Nantais2016,papovich18_smf,vanderBurg20}, in the sense that dense environments host a larger fraction of quenched galaxies. Nevertheless, such variations are second order effect in our empirical AGN-seeding model. This is shown in Figure \ref{fig:fractions}, where the predictions using the \cite{aird18_sar} specific accretion rate distribution model that includes star-formation dependence and those based on the \cite{age17_sar} specific accretion rate distribution (no SFR-dependence) are similar at $z=1.25$. 

The higher fraction of AGN in $z\ga1$ clusters compared to the field expectation in Figure \ref{fig:fractions} contradicts the results of \cite{martini13}, who report similar fractions. The field AGN fraction of \cite{martini13} is estimated from annular regions centered on the clusters with inner and outer radii of 2 and 6\,arcmin respectively. These regions may include filaments, infalling groups and generally dense structures associated with the nodes of the cosmic web where the cluster is found. They may therefore not be entirely representative of the true field. Relevant to this point are recent results by \cite{koulouridis19} who studied the radial distribution of massive galaxy clusters at $z\approx1$. They report a statistically significant overdensity of X-ray selected AGN within the infall cluster region at a project distance of $\sim2-2.5 \,R_{500}$ relative to the cluster center. This interval lies within the region used by \cite{martini13} to determine their field AGN fractions. The enhanced number of AGN found by \cite{koulouridis19} could bias high the field AGN fractions estimated by \cite{martini13}. Our findings are consistent with the higher fraction of infrared selected AGN in massive clusters of galaxies at $z\ga1$ reported by \cite{alberts16}. We caution nevertheless that the selection function of that sample is very different from that of \cite{martini13}. \cite{alberts16} identifies AGN by fitting model templates to the multi-wavelength SEDs of mid-infrared selected cluster members. 
 
\subsection{Reproducing high-$z$ AGN fractions in clusters}  
It is interesting to speculate on specific accretion rate distributions that reproduce the \cite{martini13} AGN fractions in massive clusters of galaxies at $z\approx1.25$ shown in Figure \ref{fig:fractions}. A distribution is required that produces more luminous AGN compared to the current model. There are clearly many functional forms that could achieve that. For simplicity we approach this problem by assuming a Gaussian for the specific accretion rate distribution with parameters (mean, scatter) free to vary. We caution that this problem has a broad range of non-unique solutions. This is ultimately related to the limited observational constraints that are not sufficient to break degeneracies among model parameters. There are essentially only two data points, one of which an upper limit, on the cluster AGN fraction at $z\approx1.25$ shown in Figure \ref{fig:fractions}. It is nevertheless, instructive to explore parameter combinations (mean, scatter) that yield AGN fractions consistent with the observations. In practice we assume a Gaussian specific accretion rate distribution with a given mean/scatter that is independent of stellar mass or SFR. This is applied to \textsc{UniverseMachine} galaxies (i.e. similar to Section \ref{sec:agn_galax_conec}) to produce light-cones of clusters (Section \ref{sec:lc}) on which selection effects are applied (Section \ref{sec:selec_eff}). The resulting AGN factions are then required to be within the $1\,\sigma$ uncertainty of  the \cite{martini13}  $L_X>\rm 10^{44}\, erg \, s^{-1}$ data point or larger than the lower limit for $L_X>\rm 10^{43}\, erg \, s^{-1}$ AGN in Figure \ref{fig:fractions}. The general trend is that broader distributions (larger scatter) require lower means $\lambda_{sBHAR}$ to reproduce the data. We also find that Gaussians with mean of $\lambda_{sBHAR} \ga10^{-1}$ produce too many luminous AGN for any scatter value and are therefore not allowed by the observations. Similarly a mean value of $\lambda_{sBHAR}\la10^{-3}$ produce too few luminous AGN for any scatter value and are also rejected.
 
\begin{figure*}
    \centering 
    \includegraphics[width=1\textwidth]{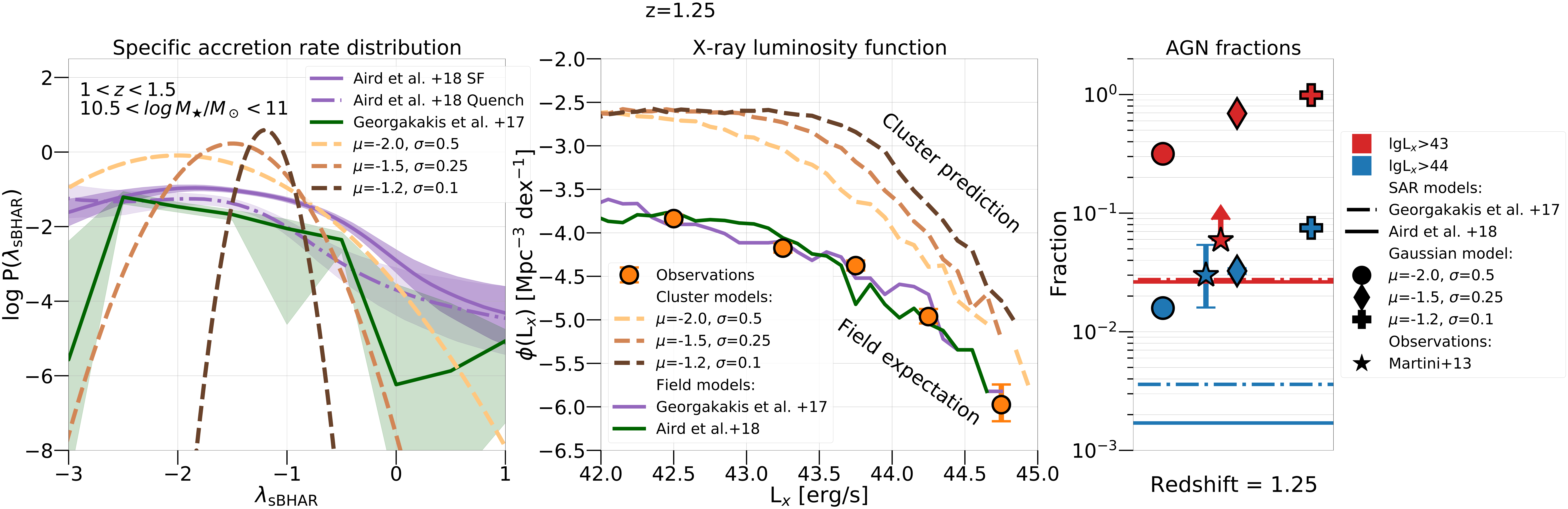}
    \caption{Comparison of Gaussian and the observationally-derived (see Section~\ref{sec:agn_galax_conec}) specific accretion rate distribution models for the cluster AGN sample described in Section~\ref{sec:selec_eff} at $z=1.25$. The left panel shows the specific accretion rate distributions that describe the probability of a galaxy hosting an AGN with specific accretion rate $\lambda_{\rm{sBHAR}}$ at $1<z<1.5$ and stellar masses $10^{10.5}<M_\star/M_\odot<10^{11}$. The observationally derived models are the purple \citep{aird18_sar} and green \citep{age17_sar} curves. Shaded region indicate 68\% confidence intervals. The yellow, light brown and dark brown curves correspond to the Gaussian distributions (see text for details) with different means and dispersions as indicated in the legend.  The middle panel compares the X-ray luminosity function for the cluster sample, predicted by the specific accretion rate distribution models (lines) with observations \citep[orange points;][]{age17_sar}. The different lines indicate the observationally derived (representative of the field population) specific accretion rate models \citep[solid, ][green and purple respectively, ]{age17_sar,aird18_sar} and the Gaussian specific accretion-rate models, proposed to match the AGN fraction in clusters at this redshift (dashed, with different colors indicating different Gaussian parameters as indicated in the legend). The right panel shows the predicted AGN fraction in clusters at $z=1.25$ following the selection effects as explained in Section~\ref{sec:selec_eff}. The stars (red or blue) are the observationally measured AGN fractions of \citet{martini09, martini13}. Different colours indicate different luminosity cut. Red is for AGN with $L_X (\rm 2-10\, keV)> 10^{43}\,erg\,s^{-1}$ and blue corresponds to $L_{X} \rm (2-10\, keV)> 10^{44}\,erg\,s^{-1}$. The predictions from the observationally derived specicific accretion-rate models in Figure~\ref{fig:fractions} are shown here with horizontal dashed-doted \citep[][]{age17_sar} and solid lines \citep[][]{aird18_sar}. The circle, diamond and plus markers correspond to the predictions of the three Gaussian specific accretion-rate models shown in the left panel with parameters as indicated in the legend.}
    \label{fig:gauss_models}
\end{figure*}

Figure~\ref{fig:gauss_models} (left panel) shows a selected number of Gaussian specific accretion rate distributions which produce AGN fractions consistent with the observations.  These distributions have more power at intermediate specific accretion rates ($10^{-2}\la \lambda_{sBHAR} \la 10^{-1}$) compared to the \cite{age17_sar} and \cite{aird18_sar} models that represent the field AGN specific accretion rate distributions. The Gaussian specific accretion rate models plotted in Figure~\ref{fig:gauss_models} make different predictions on the number of cluster AGN as a function of accretion luminosity. This is demonstrated in the middle panel of Figure~\ref{fig:gauss_models}, which plots the predicted cluster AGN X-ray luminosity functions for the different Gaussian specific accretion rate distribution models. The cluster XLF prediction is different in both shape and normalisation from the field one, also plotted in Figure~\ref{fig:gauss_models}. Future observations that provide a broad luminosity baseline and sufficient AGN number statistics can help constrain the models by e.g. directly measuring the XLF as a function of environment. We have also confirmed that even if massive clusters in the \textsc{UniverseMachine} box are assigned the higher normalisation XLFs shown in Figure~\ref{fig:gauss_models}, the total AGN XLF averaged across environments is consistent with observations. This is because massive clusters are rare as a result of the shape of the Halo Mass Function and therefore have a minor contribution to the mean XLF of the \textsc{UniverseMachine} box.

Finally, the difference between field and cluster specific accretion rates plotted in the left panel of Figure~\ref{fig:gauss_models} can be the result of varying black hole Eddington ratio distributions and/or duty cycles as a function of environment. These are the physical quantities that convolve to yield the specific accretion rate distribution  \citep[e.g.][]{Shankar2020,allevato21}. The analysis presented in this paper cannot identify which of the two quantities (Eddington ratio or duty cycle) is primary driving the differences in the AGN fraction between field and cluster environments. Further work is needed to address this issue and associate the observed differences to physical quantities directly related to the accretion flow onto the supermassive black hole.

\subsection{Evolution of AGN incidence in cluster vs. field}
Contrary to the results at $z=1.25$ discussed above, Figure \ref{fig:fractions} shows that at lower redshift, $z=0.75$, the fraction of AGN in massive clusters is consistent with the model predictions (i.e. field expectation, see Section~\ref{sec:agn_galax_conec}) within the observational data uncertainties. This conclusion does not strongly depend on the details of the semi-empirical modelling, e.g. which specific accretion rate distribution is adopted for seeding galaxies with AGN or the type of observational selection effects (mass vs. apparent magnitude cut) applied to the mock sample. At even lower redshift, $z=0.2$ our analysis tentatively suggests a paucity of AGN in cluster galaxies compared to the field. This is mainly driven by the higher (factor 2--3) fraction of AGN with $L_X ( \rm 2 - 10 \, keV ) > 10^{43} \, erg \,s^{-1}$ predicted by the model compared to the observations. For more luminous AGN, $L_X ( \rm 2 - 10 \, keV ) > 10^{44} \, erg \,s^{-1}$, no firm conclusions can be made because the observations only provide an upper limit to the AGN fraction in clusters.   We also caution that at this luminosity cut the semi-empirical model that uses the \cite{age17_sar} specific accretion rate distribution is biased high. In this model flavor a large fraction ($\approx10$\%; see Section \ref{sec:selec_eff})  of the massive central galaxies is assigned powerful AGN. This fraction is much higher than the observed incidence of luminous AGN ($\approx 1-3$\%) among the Brightest Cluster Galaxies \cite{yiang18_bcgs}. This discrepancy is ultimately related to the stellar mass dependence of the \cite{age17_sar} empirical specific accretion rates. A stronger such dependence exists in the \cite{aird18_sar} specific accretion rate distribution. As a result this model predicts much lower  $L_X ( \rm 2 - 10 \, keV ) > 10^{44} \, erg \,s^{-1}$ AGN fractions in $z=0.2$ clusters.

Overall the evidence above points to a differential redshift evolution of the incidence of AGN in clusters relative to the field. Massive structures at $z\ga1$ are found to be more efficient in triggering accretion events onto supermassive black holes compared to less dense regions. Such an environmental dependence is not present at $z\approx0.75$ and possibly inverses at low redshift, $z\approx0.2$, in the sense that clusters likely become less active regions for black hole growth compared to the field. 

\cite{Eastman2007} also find evidence for differential evolution of the AGN population as a function of environment. They compared the fraction of X-ray selected AGN in cluster of galaxies between $z\approx0.2$ and 0.6. Their analysis suggests an accelerated evolution of the AGN population in dense environments compared to the field between these redshifts. X-ray observations of individual proto-clusters at higher redshift $z\ga2$ find that AGN are more common among galaxies in these environments compare to the field \citep[][]{lehmer09,dNorth10, krishnan17}. These trends suggests that the probability of a galaxy hosting an accretion event depends on its small-scale environment, $\rm <1\,Mpc$, in agreement with our findings. 

It is also interesting that observational studies of the Halo Occupation Distribution (HOD) of AGN are broadly consistent with this picture. At low-redshift, $z\la0.2$, the HOD of X-ray AGN is proposed to have satellite fractions that increase with halo mass, albeit less rapidly than the galaxy population \citep[][]{miyaji11,allevato12}. This points to a decreasing fraction of AGN relative to galaxies with increasing halo mass, similar to our conclusions for low redshift clusters. At higher redshift however, studies of the quasi-stellar object projected correlation function find that the quasi-stellar object satellite fractions increase with halo mass similar to galaxies \citep[][]{richardson12,shen13}. The evidence above is therefore consistent with a picture whereby the fraction of AGN in massive clusters evolves with redshift.

\subsection{Physical interpretation}
Next we explore different physical mechanisms that could be responsible for the trends discussed in the previous sections. In that respect it is interesting that the redshift evolution of the fraction of AGN in clusters relative to the field bears similarities to the star-formation properties of cluster member galaxies as a function of cosmic time. In the local Universe clusters are dominated by quiescent galaxies and their integrated SFRs are significantly lower than the field expectation \citep[e.g.][]{chung2011}. This changes however toward higher redshift. The total cluster SFR normalised to halo mass increases with lookback time and possibly catches up with the field at $z>1$ \cite[e.g.][]{Webb2013, Popesso2015, alberts16}. This is also accompanied by an inversion of the SFR vs. density relation \citep{Butcher_Oelmer1984}, whereby cluster cores at $z\ga1$ host large fractions of actively star-forming galaxy populations \cite[e.g.][]{Elbaz2007, alberts16}. The trends above could be explained by the increasing fraction of the cold gas content of galaxies toward higher redshift \citep[e.g.][]{Tacconi2010, Saintonge2013, Santini2014, Gobat2020} and the increasing fraction of galaxy interactions in overdense regions compared to the field. However, it is still unclear if cluster member galaxies at $z\ga1$ are as gas rich as their field counterparts. Stacking analysis at millimeter wavelengths using ALMA continuum observations find that star-forming galaxies in massive clusters at $z\ga1$ are on average significantly more deficient in molecular gas relative to the field \citep{Alberts2022}.  This is at odds with CO emission-line observations of galaxies in clusters that typically detect molecular gas fractions comparable or higher than the field \citep[][]{Noble2019, Williams2022}. Despite this discrepancy the general picture emerging from these studies is that star-formation in massive clusters is associated with infalling galaxies during their first passage through the dense region that either manage to retain/replenish their molecular gas  \citep[e.g.][]{kotecha22, vulcani18} or have been largely stripped but continue to consume any remaining gas by forming stars before being rapidly quenched \citep[e.g. ``delay then rapid'' scenario][]{wetzel13} . 

It is therefore possible that AGN in clusters are also associated with galaxies that fall for the first time into the deep potential well of the overdensity. Relevant to this point is the discovery of an excess of X-ray AGN at the outskirts (2--3\,$R_{500}$) of massive clusters ($M_{500} > 10^{14}\, M_\odot$) at $z\approx1$ \citep{koulouridis19}. This finding points to a direct link between the growth of supermassive black holes and the infall region of massive structures in the cosmic web. Therefore the strong evolution of the AGN fraction in clusters may be linked to the increasing molecular gas content with increasing redshift of the galaxies that fall into massive halos. It is then interesting to speculate on the physical mechanisms that could lead to the {\it differential} evolution of the AGN fraction in clusters relative to the field{\bf, possible processes include:}
\begin{enumerate}
    \item Galaxy interactions. These are expected to be more common in dense environments and the outskirts of clusters. Moreover numerical simulations show that the merging rate of dark matter halos as well as the accretion rate onto them increases with redshift  \citep[e.g.][]{Gottloeber2001, Fakhouri2008, McBride2009}. Therefore the flow of galaxies from the cosmic web onto massive halos is expected to be higher at earlier times therefore leading to enhanced galaxy interaction rates. The Semi-Analytic Model of \cite{Gatti2016} suggests that galaxy interactions dominate relatively luminous AGN $L_X\ga10^{43} \rm \, erg \, s^{-1}$ in massive halos ($\ga 10^{14}\,M_\odot$) and low/intermediate redshift, $z\la1$. Instead, internal processes, such as disk instabilities, become important for lower luminosity AGN in dense environments. At much higher redshift, $z>2$, the relative contribution of galaxy interactions and internal processes in triggering AGN inverses, with disk instabilities dominating in massive halos. This is because of the higher gas and disk fraction of galaxies at these higher redshift \citep{Gatti2016}.

    \item Ram-pressure. Interaction between the Intra-Cluster Medium (ICM) and the cold gas of the galaxy, can make the latter lose angular momentum (without stripping it), hence facilitating its flow to the nuclear regions where it can accrete onto the supermassive black hole. Simulations have shown that this is possible, if the density of the ICM is not very dense (e.g. in the outskirts of clusters) \citep[e.g.][]{marshall18, Ricarte2020}. Observations of galaxies in local clusters with morphological evidence for ongoing ram-pressure stripping indeed reveal a high incidence ($\sim27-51\%$) of AGN optical emission-line signatures \citep{peluso22}. It is further expected that the ICM is more tenuous toward higher redshift and in the cluster outskirts. This is the regime where the physical conditions are more favorable for the ram-pressure to have a positive impact on AGN triggering. Such a scenario could lead to the observed trend of increasing AGN fractions in clusters relative to the field with increasing redshift. Clusters at $z\approx0.2$ have dense ICMs and the ram-pressure is sufficiently strong to strip galaxies off their gas even at the outskirts \citep[e.g.][]{zinger18, arthur19} and hence, perhaps suppress powerful AGN. Instead, at higher redshift the conditions may be appropriate for the ram-pressure to have a positive effect and boost the numbers of luminous AGN. The radial distribution of AGN within massive halos could provide further constraints this scenario \citep[e.g.][]{marshall18}.
\end{enumerate}

In addition to the processes above, simulations also highlight the potential of cosmic filaments in channeling matter and streams of cold gas deep into the potential well of massive overdensities in the Universe \citep[e.g.][]{Keres2005, Dekel2006, Keres2009}, particularly at high redshift, $z\ga1$. Galaxies that fall into clusters along such filaments are shielded from the hot ICM and can therefore retain at least part of their gas reservoirs even close to the cluster core \citep{kotecha22}. This delays the quenching of galaxies by modulating ram-pressure stripping or strangulation and allows the formation of new stars in dense cluster environments.  It can be expected that the same process also promotes black hole growth in the galaxies that accrete onto massive halos through filaments. Furthemore \cite{kotecha22} argue that this effect is more pronounced at high redshifts $z=1-3$ since at this epoch the cold flow filaments are expected to have a higher temperature contrast relative to the ICM and cluster halos are typically less massive. Such an evolution pattern is consistent with the higher fraction of AGN in massive clusters at $z\ga1$ compared to lower redshift. Additional processes (e.g. interactions, positive impact of ram-pressure) need to be invoked to explain the difference between the AGN fractions in cluster and field at $z\ga1$.

\section{Summary and conclusions}

This paper addresses the fundamental question of the role of small-scale environment ($\rm <1\,Mpc$) in triggering accretion events onto the supermassive black holes at the nuclear regions of galaxies. We tackle this issue by developing a flexible semi-empirical model that populates the dark-matter halos of cosmological N-body simulations with AGN using observational relations on the incidence of accretion events among galaxies. This zero-order assumption of the model is that the probability of accretion events are independent of the parent halo masses of galaxies, i.e. agnostic to their small-scale environment. Moreover, the observationally derived AGN incidence probabilities adopted by the model are representative of the field galaxy population, i.e. those outside massive groups or clusters. This model is used to predict the fraction of AGN in massive clusters of galaxies at different redshifts and compare against observational results by carefully taking into account observational selections effects. Any differences between model predictions and observations would point to an environmental dependence of AGN triggering mechanisms. Our main findings are

\begin{enumerate}
    \item the X-ray AGN fraction in massive clusters are larger than the model prediction at $z\sim1.25$. This points to a strong environmental dependence of AGN triggering at high redshift. Black hole accretion events are promoted in massive halos relative to the model expectation, which in turn represents the field expectation.
    
    \item the model predictions are consistent with the observed AGN fractions of interemediate redshift ($z\sim0.75$) clusters of galaxies. At this redshift it appears that massive halos do not promote or suppress AGN activity relative to the field predictions of the model.   
    
    \item at low redshift, $z\sim0.2$, the model overpredicts the fraction of $L_X(\rm 2-10\,keV) >10^{43} \, erg \, s^{-1}$ AGN in clusters compared to observations. This suggests a suppression of AGN activity in clusters relative to the field expectation of the model at $z\sim0.2$.  
    
    \item overall the points above suggest a differential redshift evolution of the AGN fraction in clusters relative to the field predictions of our semi-empirical model. 
\end{enumerate}

\noindent The observed trends above may be related to the increasing gas content of galaxies with increasing redshift, coupled with mechanisms such as galaxy interactions or ram-pressure. Both of these processes under certain conditions could promote AGN activity among galaxies that fall onto massive clusters at higher redshift.

\section*{Acknowledgements}
The authors thank the anonymous referee for their careful reading and their comments of the paper. We are grateful as well to \'Angel Ruiz Camu\~nas for his helpful comments and
advises. This project has received funding from the European Union’s Horizon 2020 research and innovation program under the Marie Skłodowska-Curie grant agreement No 860744. AL is partly supported by the PRIN MIUR 2017 prot. 20173ML3WW 002 ‘Opening the ALMA window on the cosmic evolution of gas, stars, and massive black holes’. SB acknowledges the project PGC2018-097585-B-C22, MINECO/FEDER, UE of the Spanish Ministerio de Economia, Industria y Competitividad. This research made use of Astropy,\footnote{http://www.astropy.org} a community-developed core Python package for Astronomy \citep{astropy:2013, astropy:2018}; \texttt{Colossus}\footnote{https://bdiemer.bitbucket.io/colossus/index.html} \citep{diemer18_colossus}, \texttt{Numpy}\footnote{https://numpy.org/} \citep{numpy2011}, \texttt{Scipy}\footnote{https://scipy.org/} \citep{2020SciPy-NMeth}, \texttt{Matplotlib}\footnote{https://matplotlib.org/} \citep{matplotlib2007} and \texttt{HaloMod}\footnote{\url{https://pypi.org/project/halomod/}} \citep{Murray2013, Murray2021}. For the purpose of open access, the authors have applied a CC-BY public copyright license to any author accepted manuscript version arising. \\

\section*{Data Availability}

The data products and relevant code to reproduce the results of this paper are available at \url{https://github.com/IvanMuro/agn_frac_data_release} 


\bibliographystyle{mnras}
\bibliography{biblio} 




\appendix


\bsp	
\label{lastpage}
\end{document}